\documentclass{aa}
\usepackage{amsmath}
\usepackage{amssymb}
\usepackage{txfonts}
\usepackage{natbib}
\usepackage{graphicx}

\newcommand{\drm}{\mathrm{d}}
\newcommand{\e}{\mathrm{e}}
\newcommand{\im}{\mathrm{i}}
\newcommand{\map}{M_\mathrm{ap}}
\newcommand{\kms}{km\,s$^{-1}$}
\bibpunct{(}{)}{;}{a}{}{,}

\begin{document}
\title{Weak lensing study of dark matter filaments and
  application to the binary cluster \object{A~222} and
  \object{A~223} 
  \thanks{Based on observations made at ESO/La Silla under
    program Nos. 064.L-0248, 064.O-0248, 66.A-0165, 68.A-0269.}}
\author{
  J.~P. Dietrich\inst{1}
  \and
  P. Schneider\inst{1}
  \and
  D. Clowe\inst{1}
  \and
  E. Romano-D{\'\i}az\inst{2}
  \and
  J. Kerp\inst{3}}
\institute{
  Institut f\"ur Astrophysik und Extraterrestrische
  Forschung, Universit\"at Bonn, Auf dem H\"ugel 71, 53121 Bonn,
  Germany
  \and 
  Kapteyn Astronomical Institute, University of Groningen, P.O. Box
  800, 9700 AV Groningen, The Netherlands
  \and
  Radioastronomisches Institut, Universit\"at Bonn, Auf dem H\"ugel 71, 53121 Bonn,
  Germany}
\offprints{J.~P.~Dietrich,
  \email{dietrich@astro.uni-bonn.de}}
\date{Received 24.06.2004/ Accepted 19.05.2005}
\abstract{We present a weak lensing analysis of the double cluster
  system Abell~222 and Abell~223. The lensing reconstruction shows
  evidence for a possible dark matter filament connecting both
  clusters. The case for a filamentary connection between A~222/223 is
  supported by an analysis of the galaxy density and X-ray emission
  between the clusters. Using the results of $N$-body simulations, we
  try to develop a criterion that separates this system into cluster
  and filament regions. The aim is to find a technique that allows the
  quantification of the significance of (weak lensing) filament
  candidates in close pairs of clusters. While this mostly fails, the
  aperture quadrupole statistics \citep{1997MNRAS.286..696S} shows
  some promise in this area. The cluster masses determined from weak
  lensing in this system are considerably lower than those previously
  determined from spectroscopic and X-ray observations
  \citep{2002A&A...394..395D,2000A&A...355..443P,1999ApJ...519..533D}.
  Additionally, we report the serendipitous weak lensing detection of
  a previously unknown cluster in the field of this double cluster
  system.
  }
\maketitle

\section{The cosmic web}
\label{sec:Introduction}
The theory of cosmic structure formation predicts through $N$-body
simulations that matter in the universe should be concentrated along
sheets and filaments and that clusters of galaxies form where these
intersect
\citep[e.g.][]{1983MNRAS.204..891K,1985ApJ...292..371D,1991ComPh...5..164B,1996Nature..380..603B,1999MNRAS.303..188K}.
This filamentary structure, often also dubbed ``cosmic web'', has been
seen in galaxy redshift surveys
\citep[e.g.][]{1978MNRAS.185..357J,1986ApJ...302L...1D,1986AJ.....92..250G,1989Sci...246..897G,1994ApJ...420..525V,1996ApJ...470..172S}
and more recently by \citet{2004MNRAS.351L..44B} and
\citet{2004A&A...418....7D} in the 2dF and SDSS surveys, and at higher
redshift by \citet{2001A&A...372L..57M}. Observational evidence for
the cosmic web is recently also coming from X-ray observations. E.g.
an X-ray filament between two galaxy cluster was observed by
\citet{2001ApJ...563..673T}. \citet{2003IAUS..216E.170N} and
\citet{2002A&A...394....7Z} reported possible detections of warm-hot
intergalactic medium filaments.

Because of the greatly varying mass-to-light ($M/L$) ratios between
rich clusters and groups of galaxies \citep{1999elss.conf..296T} it is
problematic to convert the measured galaxy densities to mass densities
without making further assumptions. Dynamical and X-ray measurements
of the filament mass will not yield accurate values, as filamentary
structures are probably not virialized. Weak gravitational lensing,
which is based on the measurement of shape and orientation parameters
of faint background galaxies (FBG), is a model-independent method to
determine the surface mass density of clusters and filaments. Due to
the finite ellipticities of the unlensed FBG every weak lensing mass
reconstruction is unfortunately an inherently noisy process, and the
expected surface mass density of a typical filament is too low to be
detected with current telescopes \citep{2000ApJ...530..547J}.

Cosmic web theory also predicts that the surface mass density of a
filament increases towards a cluster \citep{1996Nature..380..603B}.
Filaments connecting neighboring clusters should have surface mass
densities high enough to be detectable with weak lensing
\citep{1998wfsc.conf...61P}. Such filaments may have been detected in
several recent weak lensing studies.

\citet{1998astro-ph/9809268K} found a possible filament between two of
the three cluster in the $z=0.42$ supercluster \object{MS 0302+17}, but
the detection remains somewhat uncertain because of a possible
foreground structure overlapping the filament and possible edge
effects due to the gap between two of the camera chips lying on the
filament. Also, \citet{2004A&A...422..407G} could recently not confirm
the presence of a filament in this system.
\citet[][G02]{2002ApJ...568..141G}\defcitealias{2002ApJ...568..141G}{G02}
claim to have found a filament extending between two of the three
clusters of the \object{Abell~901/902} supercluster, but the
significance of this detection is low and subject to possible edge
effects, as again the filament is on the gap between two chips of the
camera. \citet{1998ApJ...497L..61C} reported the detection of a
filament extending from a high-redshift ($z=0.809$) cluster. Due to
the small size of the image it is unknown whether this filament
extends to a nearby cluster.

\subsection{The A~222/223 system}
\label{sec:222223-system}
A~222/223 are two Abell clusters at $z \approx 0.21$ separated by
$\sim14\arcmin$ on the sky, or $\sim2800h_{70}^{-1}$ kpc, belonging to the
\citet{1983ApJS...52..183B} photometric sample. Both clusters are rich
having Abell richness class 3 \citep{1958ApJS....3..211A}. The
Bautz-Morgan types of A~222 and A~223 are II-III and III,
respectively. While these are optically selected clusters, they have
been observed by ROSAT \citep{1997MNRAS.292..920W,1999ApJ...519..533D}
and are confirmed to be massive clusters. A~223 shows clear
sub-structure with two distinct peaks separated by $\sim 4\arcmin$
in the galaxy distribution and X-ray emission. We will refer to
these sub-clumps as A~223-S and A~223-N for the Southern and Northern
clump, respectively. A~222 is a very elliptical cluster dominated by
two bright elliptical galaxies of about the same magnitude.

\citet{2000A&A...355..443P} published a list of 53 spectra in the
field of A~222/223, 4 of them in region between the clusters
(hereafter ``intercluster region'') and at the redshift of the
clusters. Later \citet[][D02]{2002A&A...394..395D}
\defcitealias{2002A&A...394..395D}{D02} reported spectroscopy of 183
objects in the cluster field, 153 being members of the clusters or at
the cluster redshift in the intercluster region. Taking the data of
\citet{2000A&A...355..443P} and \citetalias{2002A&A...394..395D}
together, 6 galaxies at the cluster redshift are known in the
intercluster region, establishing this cluster system as a good
candidate for a filamentary connection.

\subsection{Outline}
\label{sec:outline}
This paper is organized as follows. We describe observations of the
A~222/223 system in Sect.~\ref{sec:observ-222223-syst}. Our weak
lensing analysis of this double cluster system is presented in
Sect.~\ref{sec:lensing-analysis}; we compare this to the light
(optical and X-ray) distribution in
Sect.~\ref{sec:light-distr-222223}. We find possible evidence for a
filamentary connection between the two clusters and try to develop a
statistical measure for the significance of a weak lensing detection
of a filament in Sect.~\ref{sec:quant-filam}. Our results are
discussed in Sect.~\ref{sec:discusssion}.

Throughout this paper we assume a $\Omega_\Lambda = 0.7$,
$\Omega_\mathrm{m} = 0.3$, $H_0 = 70h_{70}$~\kms~Mpc$^{-1}$
cosmology, unless otherwise indicated. We use standard lensing
notation \citep{2001Phys.Rep..340..291B} and assume that the mean
redshift of the FBG is $\overline{z}_\mathrm{FBG} = 1$.

\section{Observations of the A~222/223 system}
\label{sec:observ-222223-syst}
Imaging of the A~222/223 system was performed with the Wide Field
Imager (WFI) at the ESO/MPG 2.2 m telescope. In total, twenty 600 s
exposures were obtained in $R$-band in October 2001 centered on A~223,
eleven 900 s $R$-band exposures were taken in December 1999 centered
on A~222. The images were taken with a dithering pattern filling the
gaps between the chips in the co-added images of each field.

The $R$-band data used for the weak lensing analysis is supplemented
with three 900 s exposures in the $B$- and $V$-band centered on each
cluster taken from November 1999 to December 2000. The final $B$- and
$V$-band images have some remaining gaps and regions that are covered
by only one exposure and -- due to the dithering pattern -- do not
cover exactly the same region as the $R$-band images.

The reduction of the $R$-band image centered on A~222 is described in
detail in \citetalias{2002A&A...394..395D}. The $R$-band image
centered on A~223 was reduced in the same way. The $B$- and $V$-band
data was reduced using the GaBoDS pipeline
\citep{2003A&A...407..869S,2005astro.ph..1144E}, using
Astrometrix\footnote{\texttt{http://www.na.astro.it/$\sim$radovich/wifix.htm}}
with the USNO-A2 catalog \citep{1998USNO-A2.0....M} for the
astrometric calibration and
SWarp\footnote{\texttt{http://terapix.iap.fr/rubrique.php?id\_rubrique=49}}
for the co-addition of the individual dithered images and chips. The
$B$- and $V$-band pointings were co-added into a single frame for each
color. The PSF properties of the $R$-band pointings were so different
that they were used separately for the lensing analysis. The seeing of
the co-added $R$-band images is $0\farcs{9}$ and $0\farcs{8}$ for the
A~222 and A~223 pointings, respectively.

The $R$-band image centered on A~222 was photometrically calibrated
using Landolt standard fields and corrected for galactic extinction
\citep{1998ApJ...500..525S}, while the zero-point of the $R$-band image
centered on A~223 was fixed to match the magnitudes of objects in both
fields. Because the $B$- and $V$-band data were known to be taken under
non-photometric conditions, the red cluster sequence was identified in
a color-magnitude diagram and its color adjusted to match those
expected of elliptical galaxies at the cluster redshift, using passive
evolution and $K$-correction on the synthetic galaxy spectra of
\citet{1993ApJ...405..538B}, to account for the additional atmospheric
extinction.

Due to the greatly varying coverage of the fields, it is difficult to
give a limiting magnitude for the co-added images. The number counts
stop following a power law at 22.5--23.0~mag for the $B$- and $V$-band
images and at $24$~mag for the $R$-band images.

\section{Lensing analysis}
\label{sec:lensing-analysis}
\subsection{Lensing catalog generation}
\label{sec:lens-catal-gener}
Starting from the initial SExtractor \citep{1996A&AS..117..393B}
catalog which contains all objects with at least 3 contiguous pixels
$2\sigma$ above the background, we measured all quantities necessary
to obtain shear estimates from the KSB \citep{1995ApJ...449..460K}
algorithm. For this, we closely followed the procedure described in
\citet{2001A&A...366..717E}. 

From the KSB catalog a catalog of background galaxies used for the
weak lensing analysis was selected with the following criteria.
Objects with signal-to-noise (SNR) $<2$, Gaussian radius
$r_\mathrm{g} < 0\farcs{33}$ or $r_\mathrm{g} > 1\farcs{19}$, or
corrected ellipticity $\varepsilon>0.8$ were deleted from the sample.
Objects brighter than $R<22$ were rejected as probable foreground
objects, while all objects with $R>23$ were kept as likely background
galaxies. Objects between $22<R<23$ with colors matching those of
galaxies at redshift $z<0.5$, $-0.23 < (V-R) - 0.56 \times (B-V) <
0.67$, $0.5 < B-V < 1.6$ were not used for the lensing catalog.
Objects not detected in the $B$-band image were kept if $V-R> 1.0$.
The final catalog has 25940 galaxies, or 13.5~galaxies~arcmin$^{-2}$,
without accounting for the area lost to masked reflection rings,
diffraction spikes, and tidal tails.

\begin{figure}
  \resizebox{\hsize}{!}{\includegraphics{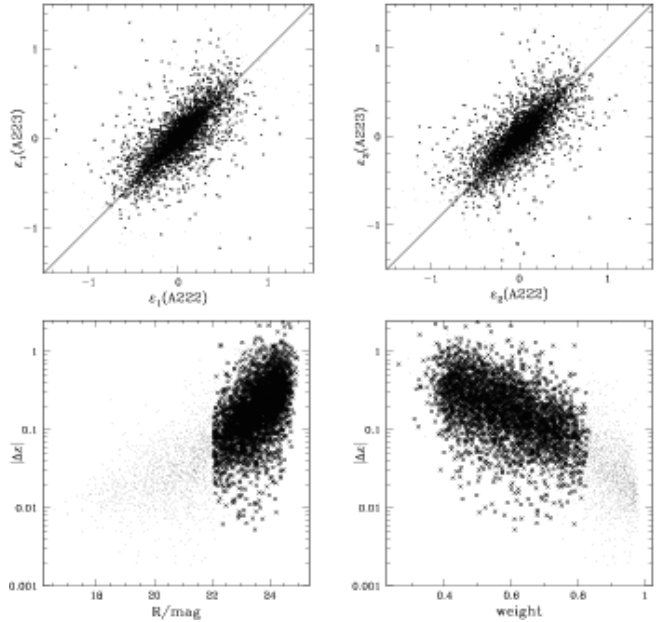}}
  \caption{Reliability tests of the shear estimates of objects observed in
    the overlapping region of the two $R$-band pointings. All objects
    with shear estimates are plotted as light dots, objects surviving
    our various selection criteria, detailed in the text, are plotted
    as crosses. \emph{Top left}: Scatter plot of the $\varepsilon_1$
    component estimates from the exposures centered on A~222 versus
    the one centered on A~223. The diagonal line is not a fit but only
    represents the ideal relation. Although we delete objects with
    corrected ellipticity $|\varepsilon| > 0.8$ from our final lensing
    catalog, some objects with $|\varepsilon| > 0.8$ are marked with
    crosses in this plot. This is because the \emph{mean} ellipticity
    of their two measurements, which we employ in our selection and
    lensing analysis is below the chosen cut-off level. These objects
    are strongly down-weighted and their exclusion would not lead to
    significant differences in the lensing analysis. \emph{Top
      right}: Same for the $\varepsilon_2$ component. \emph{Bottom
      left}: Dependence of the absolute value of differences of the
    shear estimators $|\Delta \varepsilon|$ on the apparent magnitude
    of the object. As expected, fainter objects have less reliable
    shear estimates. \emph{Bottom right}: This panels shows the
    correlation between $|\Delta \varepsilon|$ and the weighting
    scheme we employed. Objects with higher weights have more
    reliable shear estimates.}
  \label{fig:e-comp}
\end{figure}
The large overlap between the two $R$-band images allows us to test
the reliability of the shear estimates and the validity of the
weighting scheme we will employ in the lensing analysis. We perform
these tests separately for the set of all objects found in the
unmasked part of the overlap region, and the set of objects left after
performing the various cuts described in the previous paragraph. The
top panels of Fig.~\ref{fig:e-comp} show a comparison of the shear
estimates of objects observed in the overlap region of the two
$R$-band pointings. Overall, the two independent shear estimates agree
but show a broad scatter around the ideal relation. For the set of all
galaxies, we find that the mean of the differences between the two
measurements is $-0.01$ for the $\varepsilon_1$ component and $0.00$
for the $\varepsilon_2$ component. The standard deviation is $0.20$ in
each component. \citet{2003A&A...410...45E} found an $rms$ scatter of
$0.16$ between two different lensing analyses of their data. Our value
seems to indicate that the additional scatter introduced by the
independent observations is small compared to the uncertainties
intrinsic to the shear estimation procedure. The bottom left panel of
Fig.~\ref{fig:e-comp} shows the dependence of the absolute values of
differences of the shear estimators $|\Delta \varepsilon|$ on the
apparent $R$-band magnitude. As one expects, the reliability of the
shear estimates drops dramatically for fainter objects. Because these
are the objects we keep in our lensing catalog the $rms$ scatter
between the two shear estimates increases to $0.25$ per component for
the galaxies kept in our lensing catalog. The mean for the set of
galaxies in our lensing catalog stays almost unchanged at $0.01$ in
both components, showing that, while the shear estimates become
noisier, no systematic differences between both images are present.

In the lensing reconstruction and the aperture mass maps
\citep{1996MNRAS.283..837S} we will assign a weight to each shear
estimator. The weight is computed by
\begin{eqnarray}
  \label{eq:11}
  w = \left(\sigma_{\varepsilon_\mathrm{2D}}^2 +
    \sigma_\mathrm{g}^2\right)^{-1}\; ,  
\end{eqnarray}
where $\sigma_{\varepsilon_\mathrm{2D}}$ is the intrinsic 2-d
ellipticity dispersion and $\sigma_\mathrm{g}$ is the error estimate
of the initial ellipticity measurement of the galaxy. We set
$\sigma_{\varepsilon_\mathrm{2D}} = 0.38$ which is typically found in
ground-based weak lensing observations (T. Erben, private
communication; e.g. \citet{2001A&A...379..384C} who find a value of
$\sigma_{\varepsilon_\mathrm{2D}}=0.42$). $\sigma_\mathrm{g}$ is
computed from the uncertainty of the measurement of the quadrupole
moment of the galaxy in the image. While both quantities are not
independent -- $\sigma_{\varepsilon_\mathrm{2D}}$ is of course increased
by higher errors in the initial ellipticity measurement -- their
relation is very complex and not readily quantifiable in the KSB
algorithm. As a consequence, galaxies with low $\sigma_\mathrm{g}$
probably receive less weight than they should in an ideal weighting
scheme. The large overlap and the high number of objects detected in
both frames would enable us to study the shear estimation procedure in
more detail and probably find a better weighting scheme than the one
used in this work. This is, however, beyond the scope of this paper.

The bottom right panel of
Fig.~\ref{fig:e-comp} shows the correlation between our weights
(normalized to be $\le 1$) and $|\Delta \varepsilon|$. This verifies
that galaxies with more reliable shear estimates have a higher weight
in the generation of the lensing maps, although the large variations
in $|\Delta \varepsilon|$ only correspond to small relative changes in
the weight $w$. The shear estimates with the highest weight, not part
of our lensing catalog, are those which we reject as probable
foreground objects because they are too bright. 

\subsection{Weak lensing reconstruction}
\label{sec:weak-lens-reconstr}
\begin{figure*}
  \resizebox{\hsize}{!}{\includegraphics{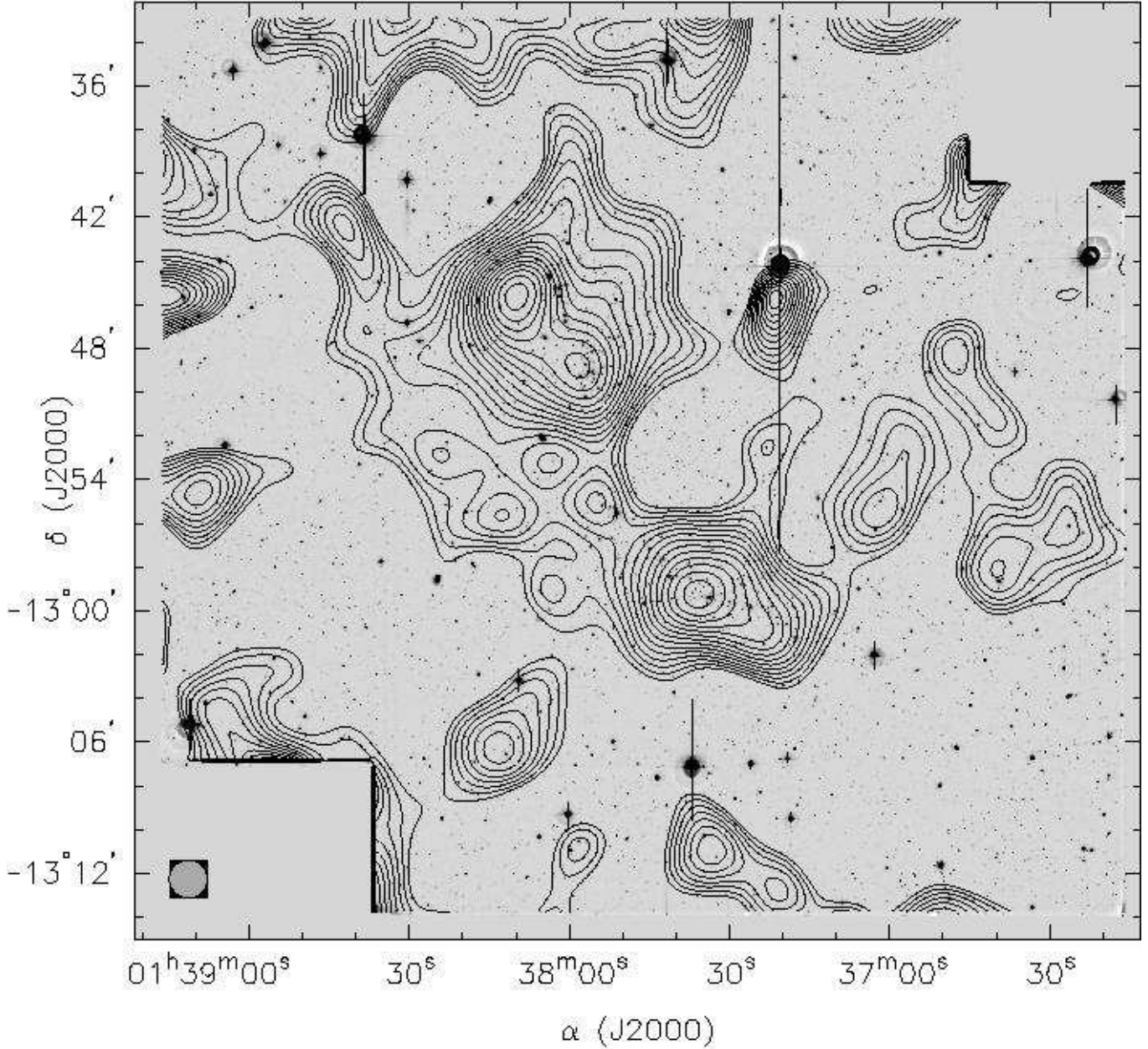}}
  \caption{Weak lensing surface mass density contours overlaid on
    the $R$-band mosaic observed with the Wide-Field imager at the
    ESO/MPG-2.2m telescope. The shear field was smoothed with a
    $\sigma=1\farcm{75}$ Gaussian, corresponding to the diameter of
    the circle at the lower left corner. Each contour represents an
    increase in $\kappa$ of 0.005 ($\sim 1.6\times
    10^{13}~h_{70}~M_{\sun}$~Mpc$^{-2}$, assuming
    $\overline{z}_\mathrm{FBG} = 1$) above the mean $\kappa$ at the
    edge of the field.}
  \label{fig:A222+3.rec}
\end{figure*}
Based on the lensing catalog described in the previous section, the
weak lensing reconstruction in Fig.~\ref{fig:A222+3.rec} was performed
using the \citet{2001A&A...374..740S} algorithm adapted to the field
geometry with a $\sigma=1\farcm{75}$ smoothing scale on a $214 \times
200$ points grid. 

Both clusters are well detected in the reconstruction, the two
components of A~223 are clearly visible, and the elliptical appearance
of A~222 is present in the surface-mass density map. The strong mass
peak West of A~223 is most likely associated with the reflection ring
around the bright $V=7.98$~mag star at that position. Although the
prominent reflection ring was masked, diffuse stray light and other
reflection features are visible, extending beyond the masked region,
well into A~223, probably being the cause of this mass peak.

The peak positions in the weak lensing reconstruction are off-set from
the brightest galaxy in A~222 and the two sub-clumps of A~223. The
centroid of the mass of A~222 is $57\arcsec$ South-East of the
brightest cluster galaxy (BCG); the mass centroids of A~223-N and
A~223-S are $86\arcsec$ and $37\arcsec$ away from the BCGs of
the respective sub-clumps.

\begin{figure}
  \resizebox{\hsize}{!}{\includegraphics{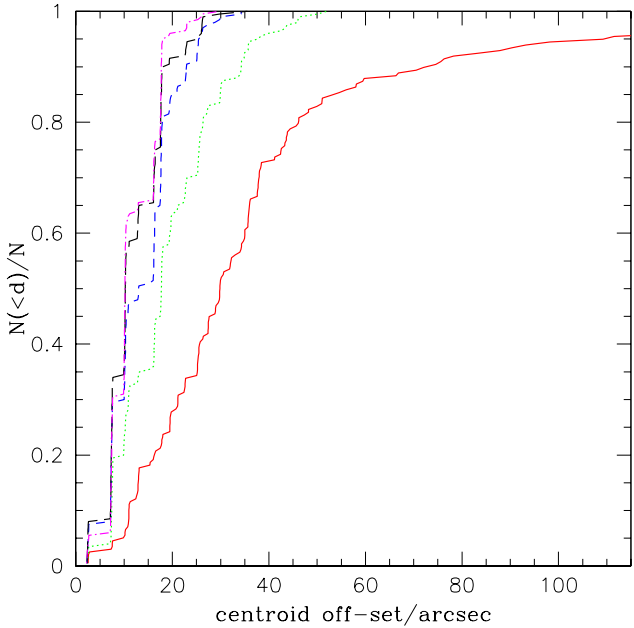}}
  \caption{Cumulative fraction of off-set of the reconstructed
    centroids from the real centroid. The curves display the
    probability of finding a reconstructed centroid of an SIS with a
    velocity dispersion of 550~\kms (continuous), 700~\kms (dotted),
    850~\kms (short dashed), 1000~\kms (long dashed), and 1150~\kms
    (dashed-dotted) from the real centroid position. The SIS was put
    at a redshift of $z=0.21$; the number density of the input catalog
    was 15~arcmin$^{-2}$.}
  \label{fig:displot}
\end{figure}
To estimate the significance of these off-sets we performed lensing
simulations with singular isothermal sphere (SIS) models of various
velocity dispersions. The SIS models were put at the cluster redshift
of $z=0.21$; catalogs with a random distribution of background
galaxies with a number density of 15~arcmin$^{-2}$, and 1-d
ellipticity dispersion of $\sigma_{\varepsilon_\mathrm{1D}} = 0.27$
were created for 200 realizations. The shear of the SIS models was
applied to the galaxy ellipticity of the catalog. Weak lensing
reconstructions based on these catalogs were performed with the
smoothing scale set $\sigma = 1\farcm{75}$ to match the smoothing of
our real data. Due to the lower SNR for the 550~\kms SIS, the
simulations yielded only 198 reliable centroid positions, while the
centroid positions of the more massive SIS could be reliably
determined in all 200 realizations. Fig.~\ref{fig:displot} shows the
cumulative fraction of reconstructed peaks found within a given
distance from the true centroid position.

These simulations show that the observed off-set of A~223-S is
compatible with the statistical noise properties of the
reconstruction. The off-set of A~222, using the the velocity
determination of the SIS models fitted below, is significant at the
2-3$\sigma$ level. The off-set of A~223-N cannot be explained with the
statistical noise of the reconstruction alone. It is, however, likely
that the observed significant off-sets are not real but linked to the
influence of the bright star and its reflection rings West of A~223.
Although objects coinciding with this reflection ring were excluded
from the catalog, the presence of a strong mass peak on the position
of the bright reflection ring is a clear indication that the shear
estimates are affected by the weaker reflection features which are too
numerous and large to be masked. It is difficult to guess how these
reflections could contribute to the observed peak shifts. We found
that varying the size of the masked region did affect the strength of
the peak on the reflection ring but left the off-sets of the cluster
peaks essentially unchanged. Still, it is noteworthy that the mass
peaks are shifted preferentially away from the star.

\begin{figure}
  \resizebox{\hsize}{!}{\includegraphics[angle=270]{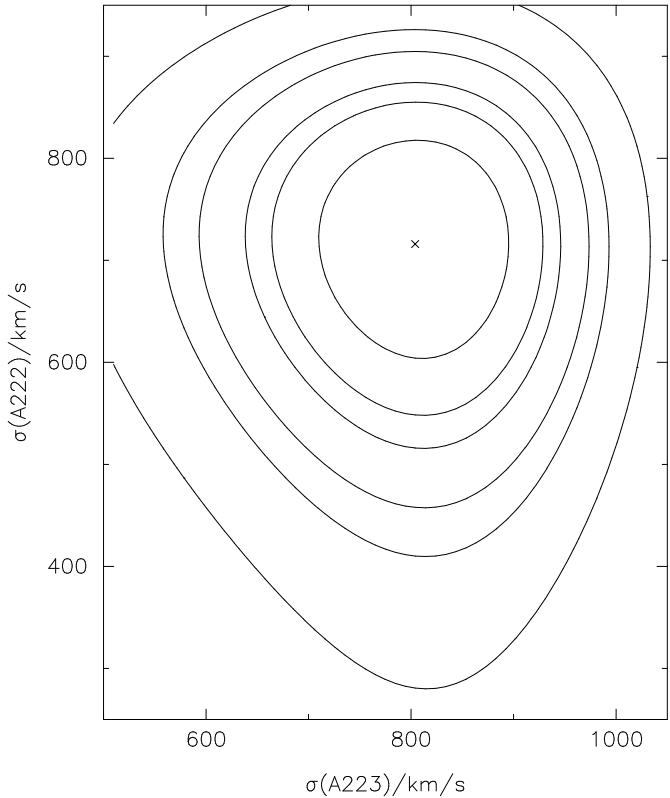}}
  \caption{Combined confidence contours for the SIS velocity
    dispersion of A~222 and A~223. The contour lines are drawn for
    $2\Delta l = \left\{ 2.3, 4.61, 6.17, 9.21, 11.8, 18.4\right\}$,
    corresponding to the 63.8\%, 90\%, 95.4\%, 99\%, 99.73\%, and
    99.99\% confidence levels, respectively, under the assumption that
    the statistics is approximately Gaussian.}
  \label{fig:com-conf}
\end{figure}

To avoid the mass-sheet degeneracy we estimate the cluster masses from
fits of parameterized models to the shear catalog. The fits were
performed minimizing the negative shear log-likelihood function
\citep{2000A&A...353...41S}
\begin{eqnarray}
  \label{eq:2}
  l_\gamma = - \sum_{i=1}^{N_\gamma} \ln
    p_\varepsilon(\varepsilon_i|g(\vec{\theta}_i))\;
\end{eqnarray}
over the $N_\gamma$ galaxy images to obtain the parameter set most
consistent with the probability distribution
$p_\varepsilon(\varepsilon_i|g(\vec{\theta}_i))$ of lensed galaxy
ellipticities. See \citet{2000A&A...353...41S} for a detailed
discussion of this maximum likelihood method for parameterized models.
We fitted more than one mass profile simultaneously. Compared to a
single model fit, this reduces the influence of the other cluster on
the fitting procedure and result. Galaxies within distances $\theta <
\theta_\mathrm{min} = 3\arcmin$ from the centers of the models were
ignored when fitting SIS models. Assuming a typical SIS this
corresponds to roughly 10 Einstein radii and is enough to assume that
all galaxies used in the fitting procedure are in the weak lensing
regime. Ignoring galaxies close to the cluster centers also reduces
the contamination with faint cluster galaxies. As a first approach we
fit two SIS, one centered on the BCG of A~222, the other centered on
the line connecting the BCGs of the two sub-clumps of A~223. The
best-fit models in this case have velocity dispersions of
$716^{+67}_{-74}$~\kms and $804^{+59}_{-64}$~\kms, respectively. This
is considerably lower than the spectroscopic velocity dispersions of
\citetalias{2002A&A...394..395D} of $1014^{+90}_{-71}$~\kms for A~222
and $1032^{+99}_{-76}$~\kms for A~223. It is also lower than the
velocity dispersions derived from X-ray luminosities
\citep{1999ApJ...519..533D} and the $L_\mathrm{X}-\sigma$ relation of
\citet{1999ApJ...524...22W}, which are $845 - 887$~\kms and $828 -
871$~\kms, respectively \citepalias{2002A&A...394..395D}, but the
value for A~223 is consistent within the error bars. The error bars of
the individual velocity dispersion were computed from $2\Delta l$,
where the velocity dispersion of one component was kept fixed at its
best-fit value to give the errors estimate for the other component.
The two component fit has a significance of $7.9\sigma$ over a model
without mass. Joint confidence contours are displayed in
Fig.~\ref{fig:com-conf}.
A three component model with an SIS centered on each BCG has a lower
significance over a zero mass model than the two SIS model and does
not fit the data better. 

Because both clusters are elliptical and the masses determined from
SIS fits differ strongly from those derived by \citetalias{2002A&A...394..395D}, one might assume
that fitting elliptical mass profiles yields a more accurate estimate
of the cluster mass. To test this, we fitted singular isothermal
ellipse models to the clusters. It turned out that the 6 parameter fit
necessary to model both cluster simultaneously was very poorly
constrained and the fit procedure was not able to reproduce the
orientation of the clusters. Results strongly depended on the initial
values chosen for the minimization routines.

\begin{figure*}
  \resizebox{\hsize}{!}{\includegraphics{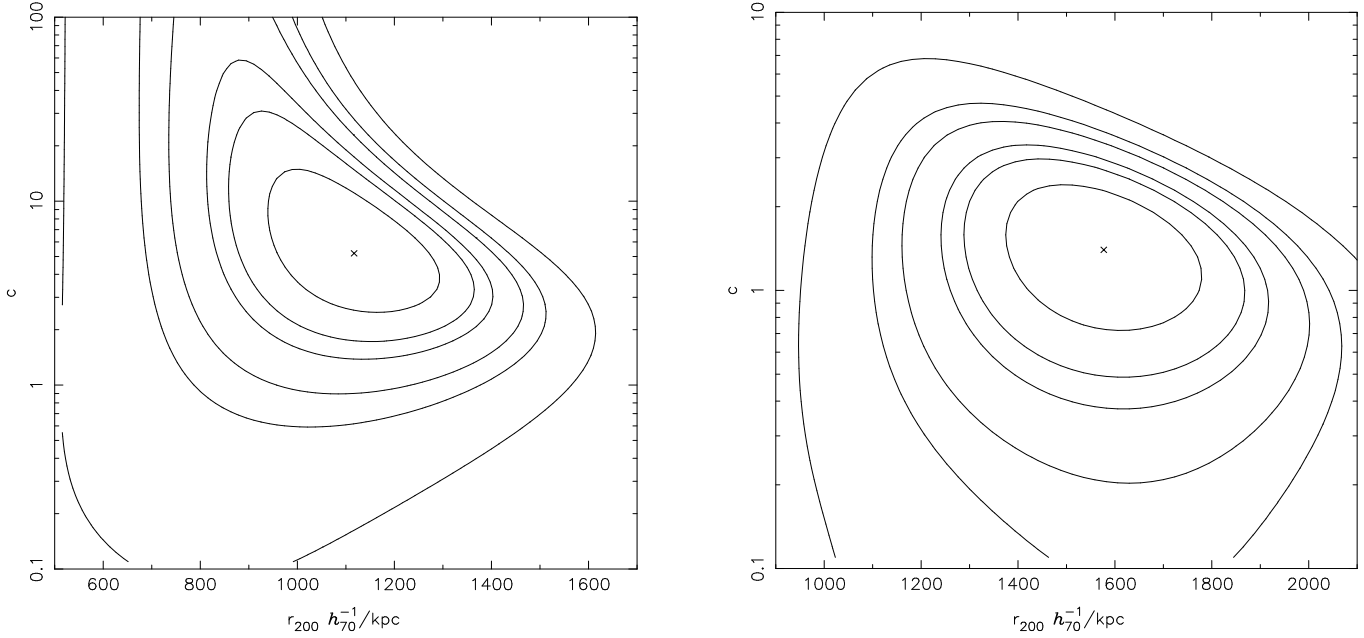}}
  \caption{Confidence contours for the best-fit NFW parameters for
    A~222 (\emph{left panel}) and A~223 (\emph{right panel}). The
    contours are drawn at the same levels as in Fig.~\ref{fig:com-conf}.}
  \label{fig:nfw-conf}
\end{figure*}
The best-fit NFW models have $r_{200} =
1276^{+102}_{-121}h_{70}^{-1}$~kpc, $c = 3.4$ and $r_{200} =
1546^{+145}_{-151}h_{70}^{-1}$~kpc, $c = 1.2$ for A~222 and A~223,
respectively, excluding shear information at distances $\theta <
1\farcm{5}$ from the cluster centers. The NFW models have a
significance of $5.2\sigma$ over a model with no mass.
Fig.~\ref{fig:nfw-conf} shows confidence contours for the NFW fits to
the individual clusters, computed from $2\Delta l$, while keeping the
best-fit parameters for the other cluster fixed. We summarize the
derived cluster properties in Tab.~\ref{tab:clus-summary}.
\begin{table}
  \centering
  \caption{Summary of the cluster properties derived from spectroscopy, X-ray, and weak lensing observations.}
  \begin{tabular}{lrr}\hline\hline
    & A~222 & A223\\\hline
    $\sigma_{\mathrm{vir}}$/(km~s$^{-1}$) & $1014^{+90}_{-71}$ &
    $1032^{+99}_{-76}$\\
    $\sigma(L_{\mathrm{X}})$/(km~s$^{-1}$) & 845--887 & 828--871\\
    $\sigma_{\mathrm{SIS}}$/(km~s$^{-1}$) & $716^{+67}_{-74}$ & $804^{+59}_{-64}$\\
    $r_{200}$/(kpc~$h_{70}^{-1}$) & $1276^{+102}_{-121}$ & $1546^{+145}_{-151}$ \\\hline
  \end{tabular}
  \label{tab:clus-summary}
\end{table}

As we see from the left panel in Fig.~\ref{fig:nfw-conf}, it can be
difficult to obtain reliable concentration parameters from weak
lensing data. The reason is that the shear signal is mostly governed
by the total mass inside a radius around the mass center. Only in the
cluster center the shear profile carries significant information about
the concentration parameter. For example if we set
$\theta_\mathrm{min} = 3\arcmin$ -- like we did for fitting SIS models
-- in the minimization procedure, $M_{200}$ remains essentially
unchanged while the concentration factor can increase dramatically.
The best fit parameters for A~222 in this case are $r_{200} = 1238
h_{70}^{-1}$~kpc, $c=7.8$. If we choose the radius
$\theta_\mathrm{min}$ inside which we ignore galaxies too big, typical
values for the scale radius $r_\mathrm{s} = r_{200}/c$ are contained
within this radius. $c$ is then essentially unconstrained, i.e. the
minimization procedure cannot anymore distinguish between a normal
cluster profile and a point mass of essentially the same mass.

The situation is different for A~223. The two sub-clumps are separated
by $\sim 4\arcmin$. This means that even ignoring shear information
within the larger $\theta_\mathrm{min} = 3\arcmin$ radius, the outer
slopes of the sub-clumps are outside $\theta_\mathrm{min}$ and the
determination of the concentration parameter gives a tight upper bound
and does not change as dramatically when the minimization is performed
only with galaxies further away from the cluster center as is the case
in A~222. Because the shear outside $\theta_\mathrm{min}$ is effectively
that of an averaged mass profile inside $\theta_\mathrm{min}$ the
measured concentration parameter is very low.

The projected cluster separation is marginally smaller than the sum of
the virial radii ($r_{\mathrm{vir}} \sim r_{200}$) derived from the
shear analysis; $r_{200}(\mathrm{A~222})+r_{200}(\mathrm{A~223}) =
2822^{+177}_{-193} h_{70}^{-1}\,\mathrm{kpc}$. We have to emphasize
that this is only the \emph{projected} separation.
\citetalias{2002A&A...394..395D} found redshifts of
$z=0.2126\pm0.0008$ for A~222 and $z=0.2079\pm0.0008$ for A~223.
Assuming that both clusters participate in the Hubble flow with no
peculiar velocity, this redshift difference of $\Delta z =
0.005\pm0.001$ translates to a physical separation along the line of
sight of $(15\pm3) h_{70}^{-1}$~Mpc. Presumably, part of the observed
redshift difference is due to peculiar velocities. In any case, it is
more likely that the two clusters are physically separated and the
virial radii do not overlap.

\subsection{A possible dark matter filament}
\label{sec:possible-dark-matter}
Also visible in the reconstruction is a bridge in the surface mass
density extending between A~222 and A~223. Although the signal of this
possible filamentary connection between the clusters is very low, the
feature is quite robust when the selection criteria of the catalog are
varied and it never disappears. Variations on the selection criteria
of the catalog let the filament shift a few arcminutes in the
East-West direction. The filament strength also changes but on closer
inspection this can be attributed almost entirely to variations in the
mass-sheet degeneracy, which is fixed by setting the mean $\kappa$ at
the edge of the field to zero. Although the field is big enough to
assume that the clusters have no considerable contribution to the
surface mass density at the edge of the field, this is a region where
the $\kappa$-map is dominated by noise. Small changes in the selection
criteria can change the value of the mass-sheet degeneracy by as much
as $\kappa_0=0.02$. This illustrates that a surface mass density
reconstruction is not suitable to assess the significance of
structures as weak as filaments expected from $N$-body simulations. We
try to develop methods to quantify the significance of this signal in
section~\ref{sec:quant-filam}.

\subsection{Other mass peaks}
\label{sec:other-mass-peaks}
In addition to the cluster peaks several other structures are seen in
the reconstruction in Fig.~\ref{fig:A222+3.rec}. Using the aperture
mass statistics \citep{1996MNRAS.283..837S} with a $6\farcm{4}$ filter
scale we find that the peak $\sim 13\arcmin$ SE of A~222 has a SNR of
$3.5$. This peak corresponds to a visually identified overdensity of
galaxies.
\begin{figure}
  \resizebox{\hsize}{!}{\includegraphics[angle=270]{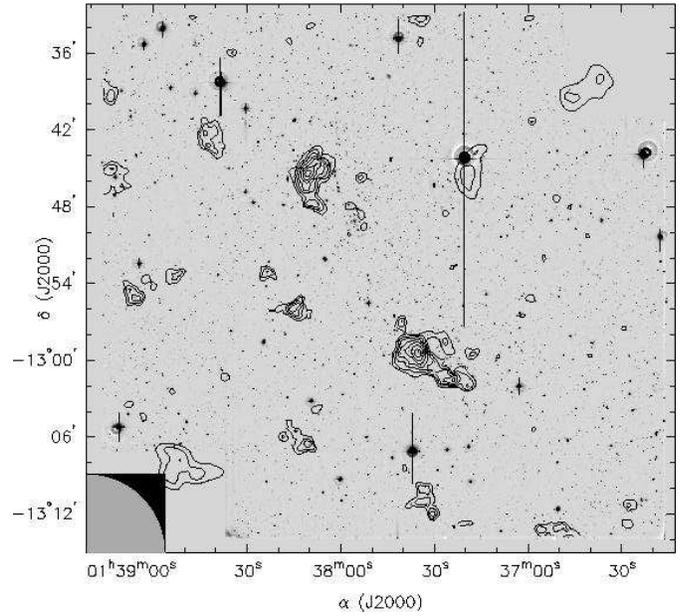}}
  \caption{Shown above are SNR contours of the aperture mass
    statistics with a $6\farcm1$ filter radius overlaid on the WFI
    $R$-band image. The lowest contour is at $2.0$, higher
    contours rise in steps $0.5$. The mass peak $\sim 13\arcmin$
    SE of A~222 has a peak SNR of $3.5$. The circle segment in
    the lower left corner has the same radius as the filter function.}
  \label{fig:a222+3.map}
\end{figure}
\begin{figure}
  \resizebox{\hsize}{!}{\includegraphics{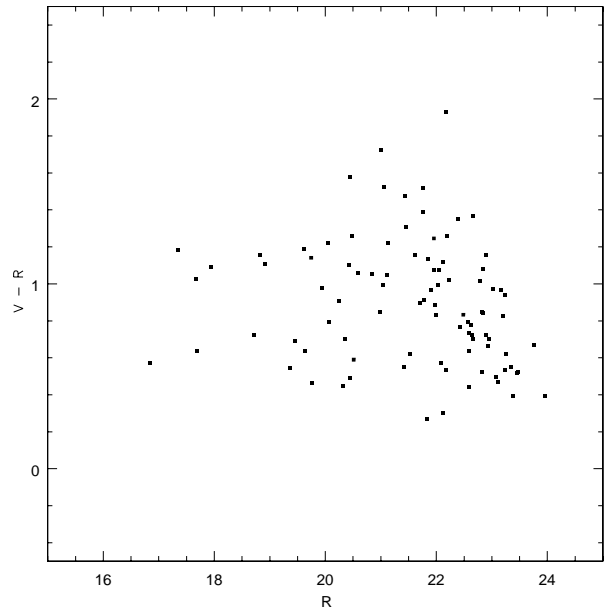}}
  \caption{Color magnitude diagram of objects around the mass peak SE
    of A~222. A possible red-cluster sequence can be seen around $V-R
    \approx 1.1$.} 
  \label{fig:blob.rcs}
\end{figure}
Fig.~\ref{fig:blob.rcs} shows a $V-R$ vs. $R$ color-magnitude diagram
of all galaxies in a box with $170\arcmin$ side length around the
brightest galaxy in this overdensity. A possible red-cluster sequence
(RCS) can be seen centered around $V-R = 1.1$, which would put this
mass concentration at a redshift of $z\sim0.4$. However, the locus of
the RCS is so poorly defined that this estimate has a considerable
uncertainty. Assuming this redshift, the best-fit SIS model has a
velocity dispersion of $728^{+101}_{-120}$\kms and a significance of
$3.2\sigma$ over a model without mass. The best-fit NFW model has
$r_{200} = 1322 h_{70}^{-1}$~kpc and $c = 3.3$ and a significance of
$2.8\sigma$ over a model with $r_{200}=0$, determined from $\delta
\chi^2$.
\begin{figure}
  \resizebox{\hsize}{!}{\includegraphics[angle=270]{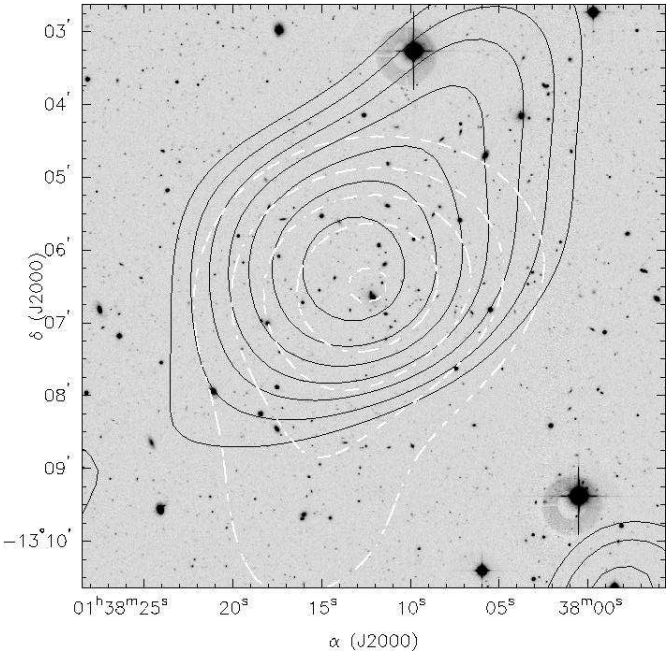}}
  \caption{Mass and light contours in the peak SE of A~222. Black
    solid lines are contours of the mass reconstruction in Fig.~\ref{fig:A222+3.rec},
    white dashed lines denote the luminosity density of galaxies with
    $1.0 < V - R < 1.2$.} 
  \label{fig:blob-contours}
\end{figure}

Fig.~\ref{fig:blob-contours} shows a comparison of surface mass and
luminosity density for this peak. Galaxies with $1.0 < V - R < 1.2$
were selected to match the tentative RCS from Fig.~\ref{fig:blob.rcs}.
The figure shows excellent agreement between the mass and light
contours, unambiguously confirming that this is a weak lensing
detection of a previously unknown cluster. The off-set between the
mass centroid and the BCG, which is located at $\alpha$=01:38:12.1,
$\delta$=$-$13:06:38.2, is $27\arcsec$ and not significant for an SIS
with a velocity dispersion of $\sim730$~\kms at a redshift of $z \sim
0.4$.

The mass concentration in the Western part of the possible filament reaches
a peak SNR of $3.6$ at a filter scale of $3\farcm{2}$. We do
not find an overdensity in the number or luminosity density of
galaxies at this position. None of the other mass peaks seen in the
reconstruction, with exception of the one on the reflection ring, is
significant in filter scales $> 2\farcm{4}$.

\section{Comparison of mass and light in A~222/223}
\label{sec:light-distr-222223}
\begin{figure*}
  \resizebox{\hsize}{!}{\includegraphics{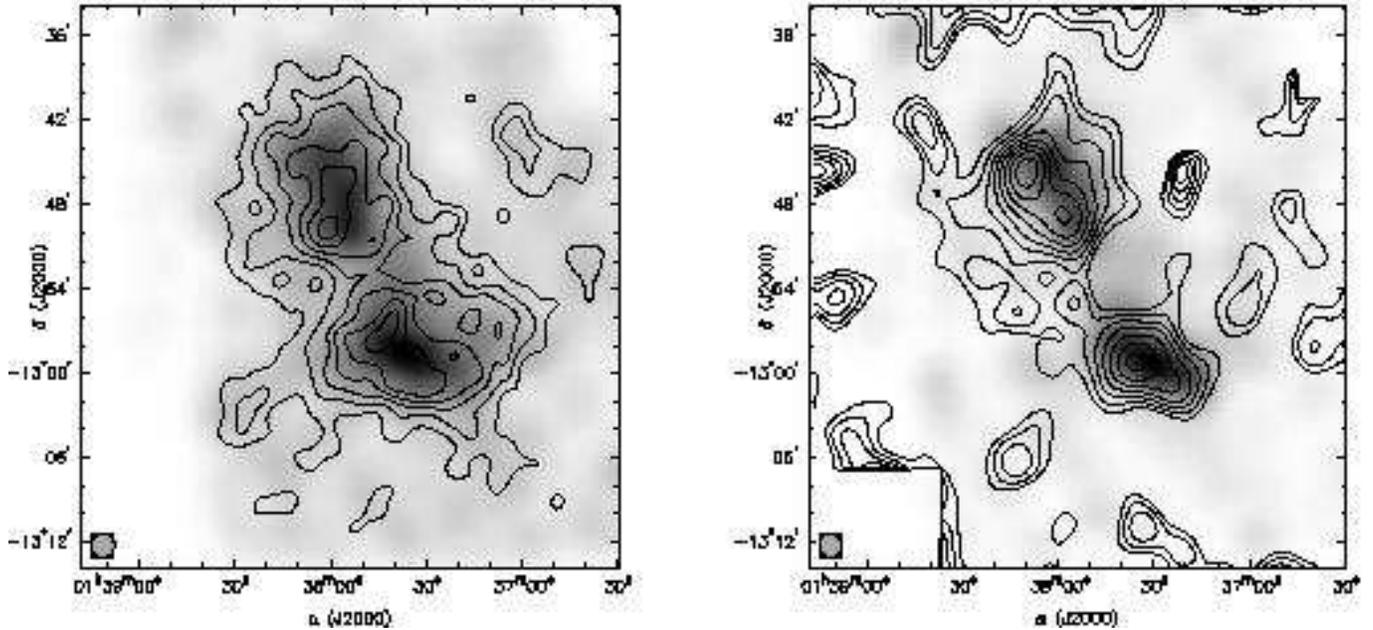}}
  \caption{Smoothed distribution of the number density (left panel)
    and the luminosity density (right panel). The smoothing was done
    with a $\sigma = 1\farcm{75}$ Gaussian to match the smoothing of
    the weak lensing reconstruction. The diameter of the
      circles at the lower left corners corresponds to the FWHM of the
    Gaussian. The contour lines in the left
    panel are significance contours starting at $5 \sigma$ and rising
    in steps of $1 \sigma$, the contour lines in the right panel are
    the surface mass density contours of the reconstruction.}
  \label{fig:nden.lumden}
\end{figure*}
Fig.~\ref{fig:nden.lumden} shows the number density and luminosity
distribution in the A~222/223 system. The left panel shows the number
density distribution of the color-selected ($0.78 < V-R < 0.98$)
early-type galaxies. The contours lines indicate the SNR determined
from bootstrap resampling the selected galaxies. It is evident that a
highly significant overdensity of early-type galaxies exists in the
intercluster region. The right panel shows a comparison between
luminosity (background gray-scale image) and surface mass density. In
general, there is good agreement between the two. We again note the
off-sets of the mass centroids from the light distribution, which we
attribute to the systematics induced by the reflection ring. The
elongation of A~222 is nicely reproduced in the reconstruction. A~222
is the dominant cluster in the luminosity density map, while in the
mass reconstruction A~223 appears to be more massive. We note,
however, that many of the bright E/S0 galaxies in A~223 escape our
color selection because they are bluer than expected for early-type
galaxies at this redshift. This indicates a high amount of merger
activity in this irregular system and most likely still collapsing
system. As we fixed the colors such that the RCS matches the expected
colors of early-type galaxies at the cluster redshift, this clearly
shows that the bright central galaxies in A~223 are bluer than
expected.

The overdensity in number and luminosity density is not aligned with
the dark matter filament candidate. We should, however, not forget
that the position of the filamentary structure is somewhat variable with
varying cuts to the lensing catalog. If the reflection features West
of A~223 are indeed responsible for shifting the centroid positions of
the massive clusters, their effect may be even stronger on such weak
features as the mass bridge seen in the reconstruction.

We estimate the cluster luminosities by measuring the $R$-band
luminosity density of all galaxies within $r_{200}$ -- as determined
from the lens models in Sect.~\ref{sec:weak-lens-reconstr} -- in
excess of the luminosity density in a circle with $5\arcmin$ radius
centered on (01:36:45.8, $-$13:07:25), which is an empty region in the
SW of our field. A~222 has a luminosity $L_{R,\,r_{200}} = (2.7 \pm
0.4) \times 10^{12}\,h^{-2}\,L_{\sun}$ and A~223 has a luminosity of
$L_{R,\,r_{200}} = (5.6 \pm 0.8) \times 10^{12}\,h^{-2}\,L_{\sun}$.
Using the mass determined from the NFW profiles, this implies rather
low $M/L$ ratios; $M/L_R =
111^{+31}_{-34}\,h_{70}\,M_{\sun}/L_{\sun}$ for A~222 and $M/L_R =
95^{+30}_{-31}\,h_{70}\,M_{\sun}/L_{\sun}$ for A~223. The mass-to-light ratios
increase to $M/L_R = 178^{+45}_{-49}\,h_{70}\,M_{\sun}/L_{\sun}$ and $M/L_R
= 131^{+30}_{-31}\,h_{70}\,M_{\sun}/L_{\sun}$ for A~222 and A~223,
respectively, if the mass estimates from the SIS models within
$r_{200}$ are used.

The X-ray satellite \textit{ROSAT} observed the pair of galaxy
clusters on 16. January 1992 using the position sensitive proportional
counter (PSPC). We extracted these data from the public \textit{ROSAT}
archive in Munich and analyzed the total integration time of 6780
seconds using the \texttt{EXSAS} software \citep{1998exsas.guide...Z}.
To avoid any confusion with diffuse soft X-ray emission and associated
photoelectric absorption towards the area of interest, we focused our
scientific interest on the upper energy limit of the PSPC detector.
Using the pulse height invariant channels 51 -- 201 (corresponding to
$0.5\,\mathrm{keV}\leq\,E\,\leq\,2.1\,\mathrm{keV}$) we calculated the
photon image and the corresponding ``exposure-map'' according to the
standard data reduction. We performed a ``local'' and a ``map'' source
detection which in total yielded 42 X-ray sources above a significance
threshold of ten.

By visual inspection, we selected some X-ray sources located close to
the diffuse X-ray emission of the intra-cluster gas and subtracted
their contribution using the \texttt{EXSAS} task
\texttt{create/bg\_image}. This task subtracts the X-ray photons of
the point sources and approximates the background intensities via a
bi-cubic spline interpolation. Finally the X-ray data were smoothed to
an angular resolution of $1\farcm{75}$ using a Gaussian smoothing
kernel.

\begin{figure*}
  \sidecaption
  \includegraphics[width=12cm]{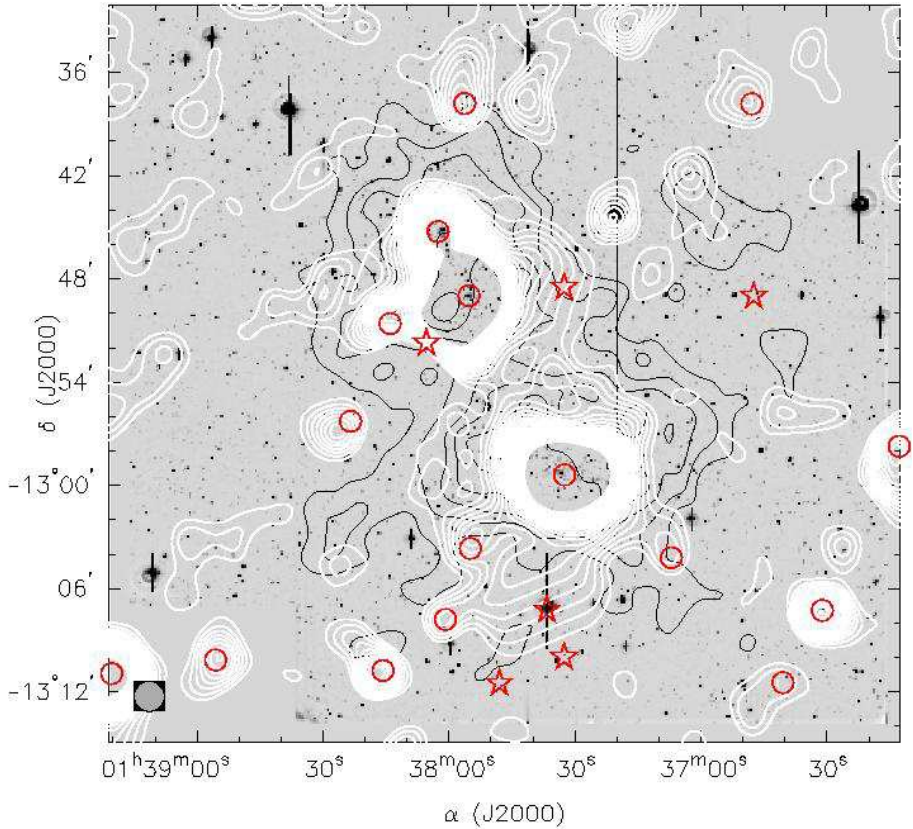}
  \caption{To the left is an overlay of X-ray contours (white lines) over
    the WFI images of our field. The contour lines start at $3\sigma$
    and increase in steps of $2\sigma$. All detected X-ray sources are
    marked; circles correspond to sources kept, while the point
    sources excised from the image are marked with stars. The X-ray
    image was smoothed with a $\sigma = 1\farcm{75}$ Gaussian,
      corresponding to the diameter of the circle at the lower left
      corner. The black contours in the background are the
    significance contours of the number density from the left panel of
    Fig.~\ref{fig:nden.lumden}. }
  \label{fig:a222+3.xray.nden}
\end{figure*}

Contours for this final image are shown in
Fig.~\ref{fig:a222+3.xray.nden}. Detected X-ray sources kept in the
final image are marked with circles; the subtracted unresolved sources
are denoted by stars. The lowest contour line is at the $3\sigma$
level. Higher contours increase in steps of $2\sigma$. Both cluster
are very well visible. As already noted by
\citet{1997MNRAS.292..920W}, A~223-S is by far the dominant sub-clump
in A~223 in X-ray.

A bridge in X-ray emission connecting both clusters is seen at the
$5\sigma$ level in this image. This possible filament is aligned with
the overdensity of the number density of color selected galaxies but
not with the filament candidate seen in the weak lensing
reconstruction.

The Eastern spur in the X-ray emission of A~223 is caused by a point
source whose removal would cut significantly into the cluster signal.
We therefore decided to keep this source. Removing it does not
influence the signal in the intercluster region. The Northern
extension of A~223 in the $\kappa$ map is blinded by the support
structure of the PSPC window in the X-ray exposure.

\section{Quantifying filaments}
\label{sec:quant-filam}
In Sect.~\ref{sec:lensing-analysis} we found a possible dark matter
filament extending between the clusters A~222 and A~223 but its
reality and significance are not immediately obvious. In this section
we try to develop a method to quantify the significance of weak
lensing filament candidate detections.

To quantify the presence of a filament and the significance of its
detection, two problems must be solved. First the fundamental question
``What is a filament?'' must be answered. How, for instance, is it
possible to discriminate between overlapping halos of two galaxy
clusters and a filament between two clusters? While in the case of
large separations this may be easy to answer intuitively, it becomes
considerably more difficult if the cluster separation is comparable to
the size of the clusters themselves (see e.g. the left panel of
Fig.~\ref{fig:nden.lumden}).

The second problem -- quantifying the significance of a filament
detection -- is rooted in the weak lensing technique. To avoid
infinite noise in the reconstruction, the shear field must be smoothed
\citep{1993ApJ...404..441K}. This leads to a strong spatial
correlation of the $rms$ error in the reconstructed $\kappa$ map,
making it difficult to interpret the error bars at any given point.
Randomizing the orientation of the faint background galaxies while
keeping their ellipticity moduli constant and performing a
reconstruction on the randomized catalog allows one to assess the
overall noise level as $\left<\kappa^2_\mathrm{rand}\right>$. While this
can be used to determine the noise level, the mass-sheet degeneracy
\citep{1995A&A...294..411S} allows us to arbitrarily scale the signal.

Statistics like the aperture mass \citep{1996MNRAS.283..837S} and
aperture multipole moments \citep{1997MNRAS.286..696S} allow the
calculation of signal-to-noise rations for limited spatial regions and
are thus well suited to quantify the presence of a structure in that
region. Hence, to quantify the presence of a structure between two
galaxy clusters, the aperture has to be chosen such that it avoids the
clusters and is limited to the filament candidate. This is of course
closely related to the first problem. We will discuss in the following
sections how aperture statistics could be used to determine the SNR of
a possible dark matter filament.

We use $N$-body simulations of close pairs of galaxy clusters to find
solutions to these problems.

\subsection{$N$-body simulations}
\label{sec:nbody}
Since we are interested in developing a method for a very particular
mass configuration, it is desirable to work with $N$-body simulations
that could mimic as closely as possible the A 222 and A 223 cluster
system. This goal can be achieved by performing \emph{constrained}
$N$-body simulations.

Constrained realizations were first explored by
\citet{1987ApJ...323L.103B}, and later presented in an elegant and
simple formalism by \citet{1991ApJ...380L...5H}. Here we follow the
approach of \citet{1996MNRAS.281...84V} of the so-called Hoffman-Ribak
algorithm for constrained field realizations of Gaussian fields. With
this approach, the constructed field obeys the imposed constraints and
replaces the unconstrained field.

The cluster-bridge-cluster system intended to simulate resembles a
quadrupolar matter distribution. It is therefore expected that
primeval tidal shear plays an important role in shaping such matter   
configuration (see \citealt{2002mtoc.conf..119V} for a review),
which must have been induced by tiny matter density fluctuations
in the primordial universe.

In moulding the observational data we have considered each one of the
observational aspects and cosmological characteristics of the system.
We have put constraints onto the initial constrained field to create
the two clusters and the bridge in the following way:

Two initial cluster seeds were sowed at the center of the simulation
box, separated by a distance $d$. We imposed constraints over the
clusters themselves like peak height (1 constraint), shape (3
constraints), orientation (3 constraints), peculiar velocity (3
constraints) and tidal shear (5 constraints).

We have performed a set of 10 realizations in a periodic $50
h^{-1}_{70}$~Mpc box, with different combinations of the mentioned 18
constraints. In all simulations we set $h_{70} = 0.7$, i.e. $H_0 =
50$~km~s$^{-1}$. All constraints were imposed over a cubic grid of
$64$ grid-cells per dimension. In all 10 realizations, all constraints
were defined on a Gaussian scale $r_g$ of $4\, h_{70}$~Mpc for both
clusters. Because we are dealing with rich clusters, we have imposed a
peak height $f_G = 3\sigma_0$, where $\sigma_0$ is the variance of the
smoothed density field ($\sigma_0(r_g) = \langle f_g f_g \rangle^{1/2}$). The
other constraints considered were: oblate clusters with axis ratios
$\lambda_2/\lambda_1 =0.9$ and $\lambda_3/\lambda_1 = 0.8$ and both
major axes aligned with each other. We have imposed a ``weak''
primordial tidal field in order to produce a realistic field around
the clusters. The stretching mode of the tidal field was aligned
along the same direction given by the major cluster axes. The
compressional mode was set perpendicular to the bridge axis. This
combination of constraints proved to be the most successful one in
reproducing (in the linear regime) the configuration presented by the
two Abell clusters.

\begin{figure}
  \resizebox{\hsize}{!}{\includegraphics{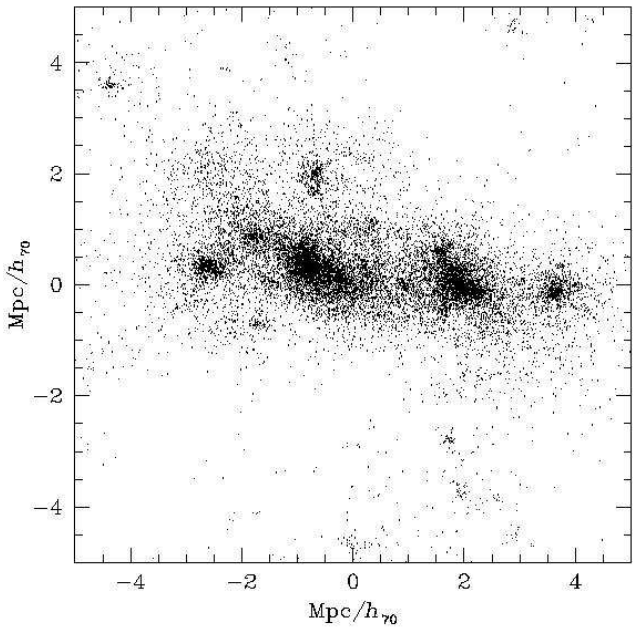}}
  \caption{Zoom in on the central
    $10\times10$ Mpc$^2$/$h_{70}^2$ of an $N$-body simulation.
    Displayed is the projection of a slice of 2.5 Mpc/$h_{70}$
    thickness.}
  \label{fig:simdata}
\end{figure}

The initial particle displacements and peculiar velocities were
assigned according to the \citet{1970A&A.....5...84Z} approximation
from the constrained initial Gaussian density field. The evolution of
the linear constrained density field into the non-linear regime was
performed by means of a standard P$^3$M code
\citep{1991ComPh...5..164B}. The number of grid-cells used to evaluate
the particle-mesh force was $128^3$, with a particle mass resolution of
$3.3 \times 10^{10} M_{\sun}$. We selected 15 time outputs in order to
follow the simulation through the non-linear regime, with a time output
at redshift $z=0.21$, to match the observed cluster redshift.
Fig.~\ref{fig:simdata} shows the most successful cluster-bridge-cluster
configuration.

\begin{figure}
  \resizebox{\hsize}{!}{\includegraphics[angle=270]{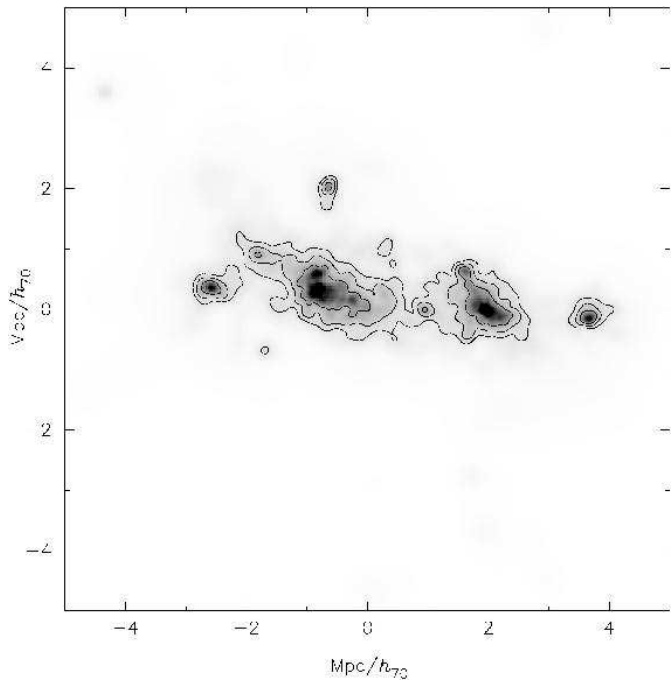}}
  \caption{Smooth density distribution of the data in the left panel
    from the adaptive kernel density estimate. The contours are at
    $\kappa = \{0.03,0.05, 0.1, 0.5\}$.}
  \label{fig:simsmooth}
\end{figure}

To estimate the underlying smooth mass distribution from the result of
the simulations, the particle distribution was smoothed using the
adaptive kernel density estimate described by
\citet{1996MNRAS.278..697P, 1993MNRAS.265..706P}. A comparison of the
simulated and smoothed mass distribution can be found in
Figs.~\ref{fig:simdata} and \ref{fig:simsmooth}.
The surface mass density of all simulations was linearly scaled such
that $\kappa_{\mathrm{max}} \simeq 1$.

We have also computed the density field by means of the Delaunay
Tessellation Field Estimator \citep{2000A&A...363L..29S}, which in
principle offers higher spatial resolution at both, dense and
underdense regions in comparison with other fixed-grid or adaptive
kernels density reconstruction procedures.

The reconstructed DTFE surface mass density maps of the
cluster-bridge-cluster system agree with those from the adaptive
kernel since in high-density regions, both methods give similar
density estimates \citep{2003A&A...403..389P}.

The lensing properties of the smoothed mass distribution were computed
on a $2048\times2048$ points grid using the \texttt{kappa2stuff}
program from Nick Kaiser's IMCAT software
package.\footnote{\texttt{http://www.ifa.hawaii.edu/$\sim$kaiser/imcat/}}
\texttt{kappa2stuff} solves the Poisson equation
\begin{eqnarray}
  \label{eq:1}
  \nabla^2 \psi(\vec{\theta}) = 2 \kappa(\vec{\theta})
\end{eqnarray}
in Fourier space on the grid by means of a Fast Fourier Transformation
(FFT) and returns -- among other quantities -- the complex shear
$\gamma(\vec{\theta})$ and the magnification $\mu(\vec{\theta})$.

For the lensing simulation, catalogs of background galaxies were
produced. Galaxies were randomly placed within a predefined area until
the specified number density was reached. To each galaxy an intrinsic
ellipticity was assigned from two Gaussian random deviates. Unless
noted otherwise all simulations have 30 galaxies/arcmin$^2$ and a
one-dimensional ellipticity dispersion of
$\sigma_{\varepsilon_\mathrm{1D}} = 0.2$.

\begin{figure}
  \resizebox{\hsize}{!}{\includegraphics[angle=270]{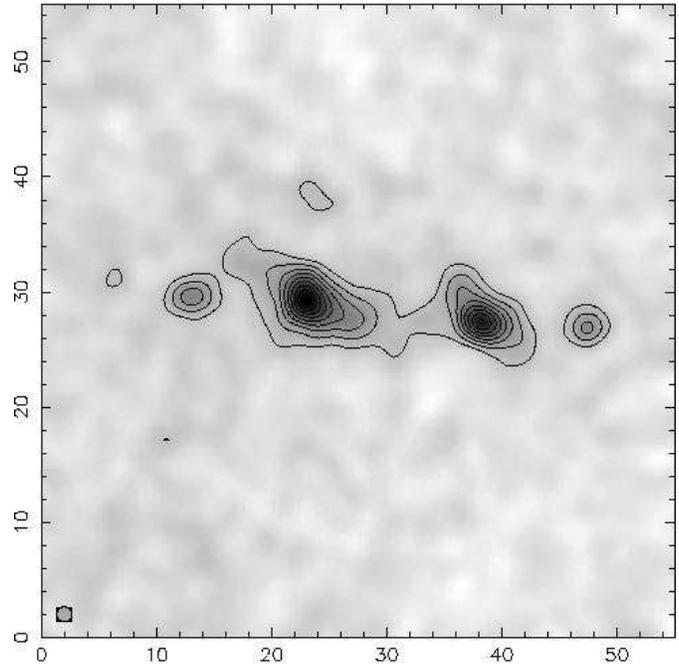}}
  \caption{Reconstruction of the mass distribution in
    Fig.~\ref{fig:simsmooth} on a $206 \times 206$ points grid.
    The scale of the axes is given in arcminutes. The contours mark an
    increase of $\kappa$ in steps of 0.025 above the mean of the edge
    of the field. The diameter of the circle in the lower left
    is equal to the FWHM of the Gaussian smoothing of the shear field.}
  \label{fig:rec.48.cluster_4.4}
\end{figure}

To test the validity of our lensing simulation we performed a mass
reconstruction of the simulation in Fig.~\ref{fig:simsmooth} using the
algorithm of \citet{2001A&A...374..740S}. The smoothing scale in this
reconstruction was set to $1\farcm{3}$. The result is shown in
Fig.~\ref{fig:rec.48.cluster_4.4}. We see that the reconstruction
successfully recovers the main properties of the density distribution;
both clusters are clearly detected, their ellipticity and orientation
agrees with that of the smoothed density field. A ``filament''
resembling that in Fig.~\ref{fig:simsmooth} is also seen. We now have
to find a way to determine the significance of its detection.

\subsection{Fitting elliptical profiles to galaxy clusters}
\label{sec:fitt-ellips-galaxy}
In a first attempt to quantify filaments, we try to fit the galaxy
clusters by elliptical mass profiles. We then define the filament as
the part of the mass distribution which is in excess of the mass
fitted by the ellipses.

The profile we used is the non-singular isothermal ellipse of
\citet{1998ApJ...495..157K}. This profile has four parameters per
cluster that need to be fitted:
\begin{itemize}
\item The axis ratio
\item The core radius
\item The Einstein radius of the corresponding singular isothermal sphere
\item The orientation of the ellipse
\end{itemize}
The central position was fixed and taken to be the peak position of
the mass reconstruction. As in Sect.~\ref{sec:weak-lens-reconstr} we
use the shear-log-likelihood function to fit more than one mass profile
simultaneously.

Various methods for multidimensional minimization are available. All
programs used for fitting either used the Down\-hill Sim\-plex or Powell's
Direction Set algorithms discussed in detail in
\citet{1992nrca.book.....P}. We could not find any systematic
differences between the results of the two methods. In general, their
results agreed quite well if the same initial values were used.

\begin{figure*}
  \resizebox{\hsize}{!}{\includegraphics{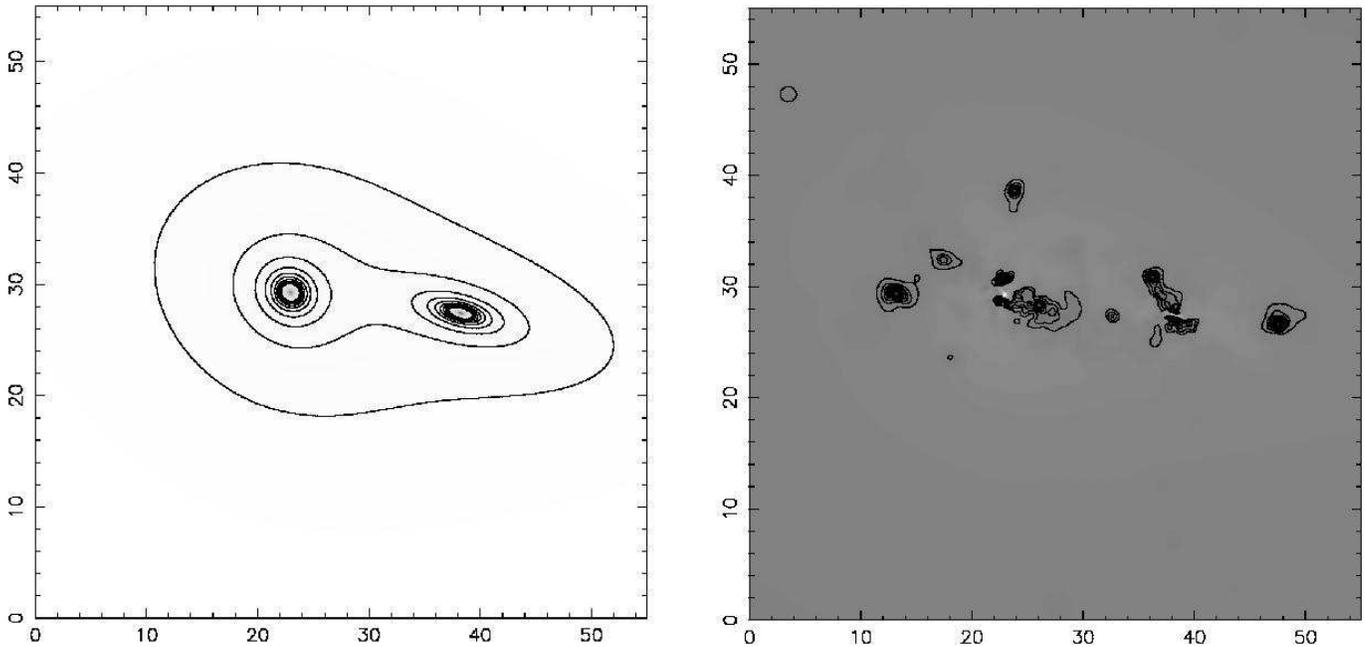}}
  \caption{\emph{Left}: Fit of two NIE profiles to the simulation in
    Fig.~\ref{fig:simsmooth}. The contour lines increase from $\kappa =
    0.025$ to $\kappa=0.25$ in steps of 0.025.
    \emph{Right}: Difference image of Fig.~\ref{fig:simsmooth} and the
    left panel. The contours are at the same level as in the left
    panel. The filament was completely subtracted. The surface mass
    density in the intercluster region is mostly negative.}
  \label{fig:niefit}
\end{figure*}
Fig.~\ref{fig:niefit} shows a fit of two non-singular isothermal
ellipse (NIE) profiles to the simulation in Fig.~\ref{fig:simsmooth}.
We see that the cluster are roughly fitted by the NIE profiles. Like
in the case of fitting SIE models to A~222/223, the ellipticity of the
original cluster is only poorly reproduced. Varying the initial values
of the minimization procedure, gives comparable best-fit values for
the Einstein and core radius. The axis ratio and orientation of the
ellipse are so strongly affected by the choice of initial values, that
they have a profound impact on the surface mass density in the
intercluster region. Especially, the orientation is only poorly
constrained. Generally, the fits overestimate the surface mass density
in the intercluster region, fitting the ``filament'' completely away.
The behavior of the fits to the simulated data confirms our experience
with fitting SIE profiles in the A~222/223 system. Letting the slope
of the density profile vary does not remedy the problem. The shear
log-likelihood function is rather sensitive to the slope of the
density profile, but the ellipticity of the clusters is still poorly
constrained.

\subsection{Using aperture multipole moments to quantify the presence of
  a filament}
\label{sec:using-apert-mult}
Aperture multipole moments (AMM) quantify the weighted surface mass
density distribution in a circular aperture. If it is possible to find
a characteristic mass distribution for filaments and express it in
terms of multipole moments, AMM can be used to quantify the presence
of a filament.
\begin{figure}
  \resizebox{\hsize}{!}{\includegraphics{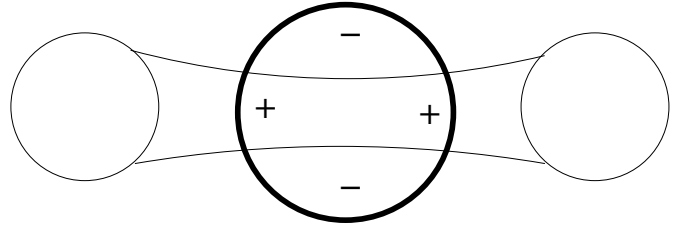}}
  \caption{Simple toy model of two galaxy clusters connected by a
    filament. A quadrupole moment is present in the aperture
    centered on the filament.}
  \label{fig:qpole_illus}
\end{figure}
\begin{figure}
  \resizebox{\hsize}{!}{\includegraphics{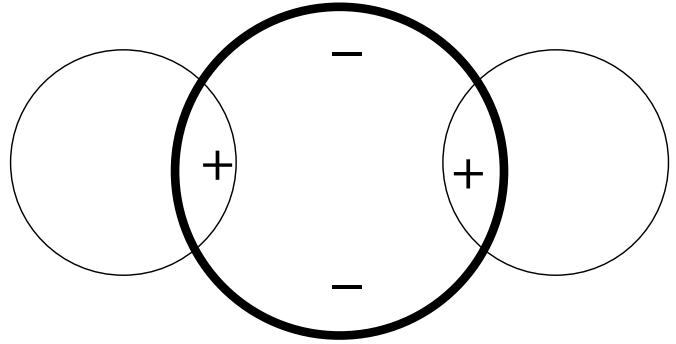}}
  \caption{Toy model of two
    galaxy clusters without a filament, illustrating why it is
    important to choose the correct size of the aperture.}
  \label{fig:qpole_illus_fake}
\end{figure} 

Fig.~\ref{fig:qpole_illus} illustrates with the help of a simple toy
model of two galaxy clusters connected by a filament why one expects
to find a quadrupole moment in an aperture centered on the filament.
Fig.~\ref{fig:qpole_illus_fake} illustrates that it is crucial not to
choose the aperture too large. If the aperture also covers the
clusters, a quadrupole moment will be measured even if no filament is
present. 

\begin{figure}
  \resizebox{\hsize}{!}{\includegraphics{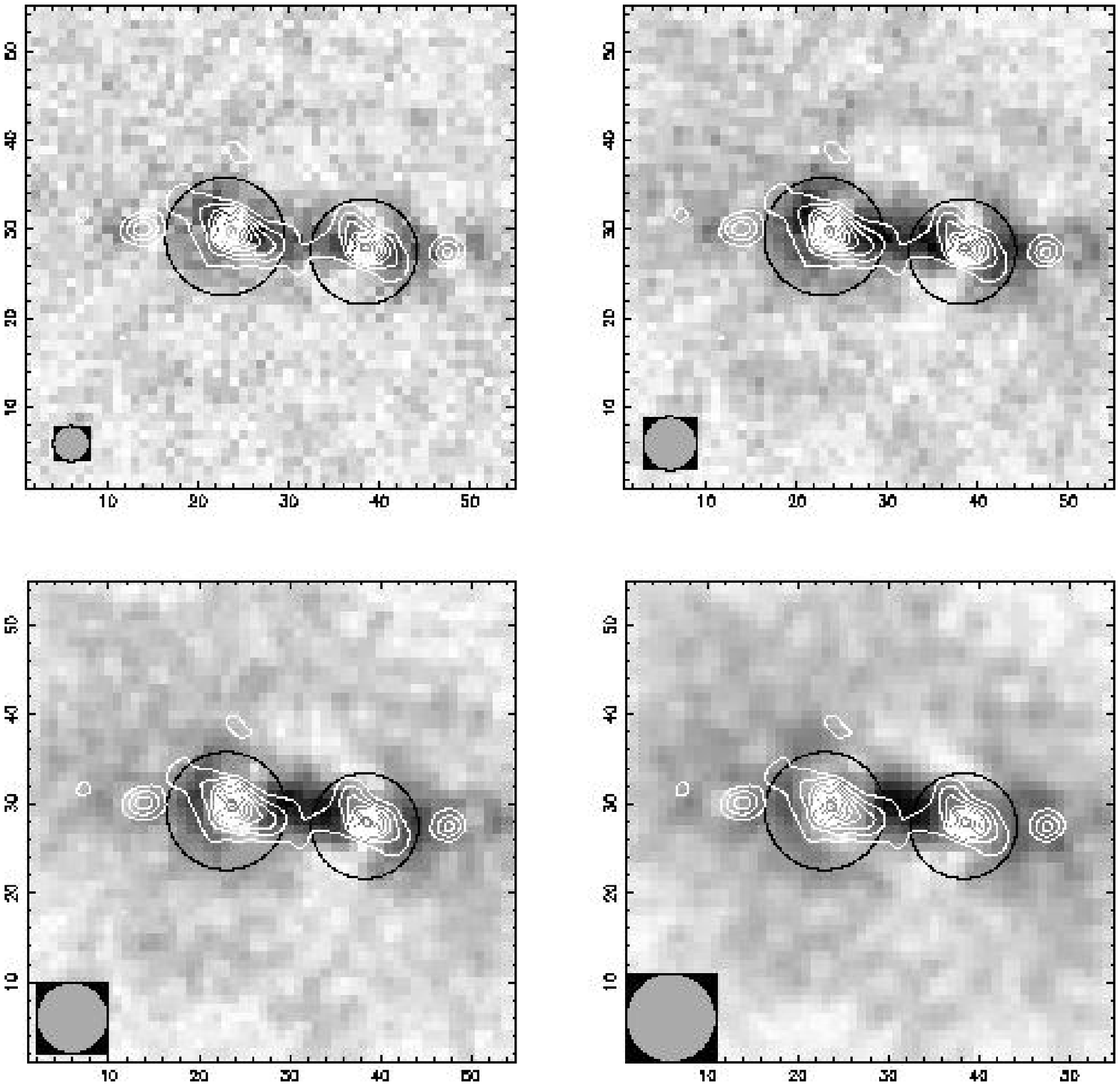}}
  \caption{Quadrupole moment of
    the simulation in Fig.~\ref{fig:simsmooth}. Overlayed are the
    contours of the mass reconstruction in
    Fig.~\ref{fig:rec.48.cluster_4.4} and large circles with radius
    $r_{200}$ as determined from the 3-dimensional simulated data.
    $|Q^{(2)}|$ was computed in an aperture with radius $\theta_\mathrm{max}
    = 2\arcmin$ (\emph{top left}), $\theta_\mathrm{max} = 3\arcmin$
    (\emph{top right}), $\theta_\mathrm{max} = 4\arcmin$ (\emph{bottom
      left}), $\theta_\mathrm{max} = 5\arcmin$ (\emph{bottom
      right}). The circles at the lower left corners have the
      same radii as the filter function in the respective panel.
    The cluster on the left has a mass of $M_{200} = 2.0 \times
    10^{14} M_{\sun}$, the one on the right of $M_{200}=1.4 \times
    10^{14} M_{\sun}$.}
  \label{fig:qpoles}
\end{figure}

We choose a weight function
\begin{eqnarray}
  \label{eq:3}
  U(\theta) = 
  \begin{cases}
    1-\left(\frac{\theta}{\theta_\mathrm{max}}\right)^2 & \theta \le
    \theta_\mathrm{max}~, \\
    0 & \text{otherwise}~,
  \end{cases}
\end{eqnarray}
to compute the aperture quadrupole moment as defined in \citet[see
also Appendix \ref{sec:multipole-moments}]{1997MNRAS.286..696S}.
While this weight function is clearly not ideal as it does not closely
follow the mass profiles of the simulated data, it is sufficient to
identify all relevant features in quadrupole moment $|Q^{(2)}|$ maps.
Fig.~\ref{fig:qpoles} shows such $|Q^{(2)}|$ maps for the simulation
in Fig.~\ref{fig:simsmooth}. In the quadrupole maps
$\theta_\mathrm{max}$ increases from 2\arcmin to 5\arcmin. The maps
were computed on $55 \times 55$ points grid, so that each grid point
is $1\arcmin \times 1\arcmin$ big. Overlayed are the contours of the
surface mass density of the reconstruction of
Fig.~\ref{fig:rec.48.cluster_4.4}.

One clearly sees that the quadrupole moment between the clusters
increases as the size of the apertures increases. This is of course
expected and due to the growing portion of the clusters in the
aperture, so that their large surface mass density dominates the mass
distribution. Fig.~\ref{fig:qpoles} also illustrates the problem of
separating clusters from a filament. The virial radii of the clusters
extend far beyond the mass contours of the clusters in most directions
and thus beyond what can be detected with weak lensing. Due to the
ellipticity of the clusters it is not obvious whether the projected
mass extending out to the virial radius is part of the cluster or
belongs to a filament. While we certainly can define that projected
mass outside the virial radii belongs to a filament, the case is not
clear for mass inside $r_{200}$. The surface mass density contours in
Fig.~\ref{fig:qpoles} seem to suggest that a signature of a filament
is present and observable with weak lensing inside the virial radius.
Because weak lensing only has a chance to detect filaments in close
pairs of clusters whose separation is comparable to the sum of their
virial radii, understanding this signature is important.

In this context the two maps in the top panel which show the smallest
overlap of the aperture with the virial radii are the most
interesting. Noteworthy in Fig.~\ref{fig:qpoles} is also that the top
panels show a quadrupole moment on a ring-like structure around each
cluster center. This is indeed to be expected for all galaxy clusters
because there is a non-vanishing quadrupole moment if the aperture is
not centered on the cluster center, but somewhere on the slope of the
mass distribution. This now raises the question how we should
distinguish the quadrupole moment present around any cluster from that
caused by a filamentary structure between the clusters.

\begin{figure}
  \resizebox{\hsize}{!}{\includegraphics{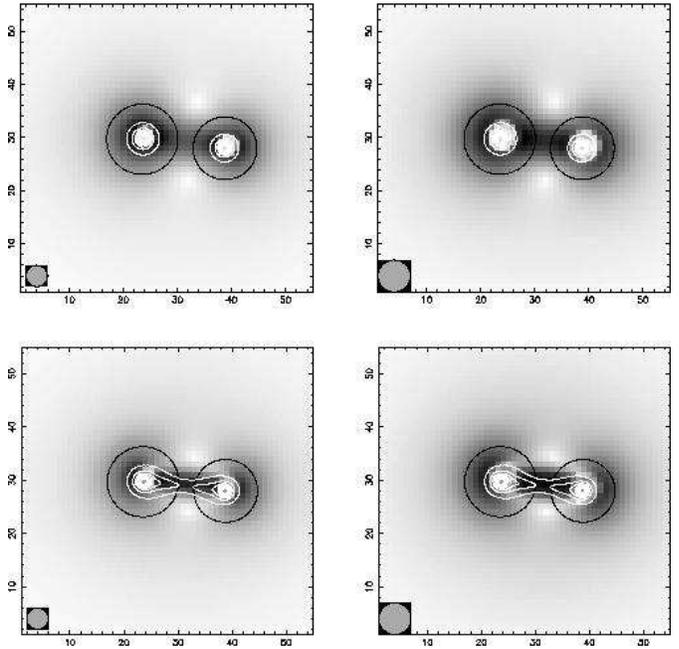}}
  \caption{$|Q^{(2)}|$ maps of toy models of two clusters. \emph{Top
      panel}: without a connecting filament. \emph{Bottom panel}: with
    a filament running between both clusters. Quadrupole moments in
    the left panel were computed in a $2\arcmin$ aperture, in the
    right panel in a $3\arcmin$ aperture. The radii of the apertures
    correspond to the radii of the circles at the lower left corners
    of the respective panel. White contours are surface mass density;
    the black circles correspond the $r_{200}$ of the clusters.}
  \label{fig:qpoles_toy}
\end{figure}

To better understand the features visible in the $|Q^{(2)}|$~maps of
the $N$-body simulations we qualitatively examine the structures of a
quadrupole map for a system of two isolated clusters and two clusters
connected by a filament using simple toy models.
Fig.~\ref{fig:qpoles_toy} shows noise-free quadrupole maps of two
truncated NFW halos \citep{2003MNRAS.340..580T} without (top panel)
and with (bottom panel) a connecting filament for the same aperture
sizes as in the top panel of Fig.~\ref{fig:qpoles}. The halo centers
and virial radii were chosen to match those in the $N$-body
simulation. To describe the filament we choose a coordinate system
such that the $x$-axis runs through the halo centers and has its
origin at the center between the clusters separated by a distance $d$
and define the following function:
\begin{eqnarray}
  \label{eq:12}
  \kappa_{\mathrm{fil}}(x, y) = k_0 \frac{\left(k_1 x^4 + k_2
      x^2\right) \left|\frac{d}{2} - x\right| 
  }{\left(y/k_3\right)^2 + 1}
  H\left(\left|\frac{d}{2}-x\right|\right)\, 
\end{eqnarray}
where $k_1 < 0$ and $k_2$ chosen such that the maxima of the fourth
order polynomial coincide with the halo centers. $H(x)$ is the
Heaviside step-function.

As in the $|Q^{(2)}|$~maps of the $N$-body simulations a qua\-dru\-pole
moment related to the slope of the clusters is visible. Already the
smallest aperture (left panel) overlaps the truncated NFW halos and
leads to a strong quadrupole moment in the intercluster region. This
quadrupole moment is, however, much weaker than it is in the presence
of a filament (bottom panel of Fig.~\ref{fig:qpoles_toy}).
Additionally, the presence of a filament is indicated by a ring
structure on which the quadrupole moment is lower. This structure
becomes more prominent with increasing filter radius. All aperture
statistics act as bandpass filters on structure comparable in size to
the filter radius. As the ring has a radius of $\sim 7\arcmin$ it is
better visible in the map generated from the larger filter. This
structure is present only in the halo-filament-halo system and not in
the halo-halo system, even for filter scales larger than those
depicted in Fig.~\ref{fig:qpoles_toy}. This structure is also visible
in the quadrupole maps of the $N$-body simulations. It is well visible
in the top right panel of Fig.~\ref{fig:qpoles} and less well visible
but still present in the top left panel of Fig.~\ref{fig:qpoles}.
Thus, the quadrupole maps clearly indicate that the measured
quadrupoles on the filament are not caused by a situation without
filament like that illustrated in Fig.~\ref{fig:qpole_illus_fake}.
Unfortunately, this ring structure, like the filament itself, is a
visual impression that does not lend itself easily to a quantitative
assessment of the presence of a filament.

These toy models and the $N$-body simulation show that the aperture
quadrupole statistics is indeed sensitive to filamentary structures.
The quadrupole moment of a halo-filament-halo system exceeds that of a
pure halo-halo system. As the measured quadrupole moment strongly
depends on the size of the aperture, choosing the appropriate aperture
is important. A decomposition of a halo-filament-halo system into halo
and filament components could be provide an objective criterion for
chosing the filter scale.

While in the (failed) attempt to separate the clusters and the
filament by fitting elliptical profiles to the clusters, the filament
was naturally defined as the surface mass density excess above the
clusters, there is no criterion in the AMM statistics that defines
cluster and filament regions. We try to develop such a criterion in
the following section.

\subsection{Defining cluster and filament regions}
\label{sec:defin-clust-filam}
\begin{figure}
  \resizebox{\hsize}{!}{\includegraphics{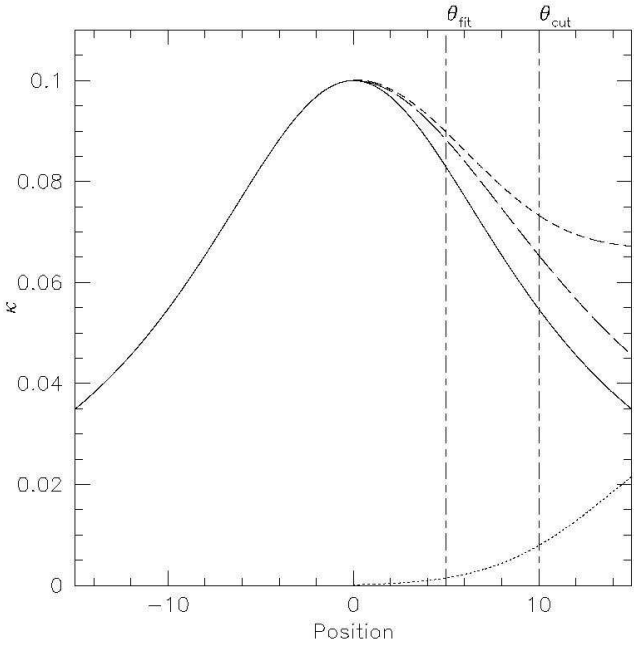}}
    \caption{Simple model of the surface mass density distribution of
      an elliptical cluster and a filamentary extension along the main
      axis of the system. The solid line is a symmetric King profile,
      the long dashed line is the same King profile stretched by a
      factor to introduce the ellipticity seen in simulation. The
      filament is modeled as a separate component (dotted line). The
      observed profile (short dashed line) is the sum of the filament
      and the stretched King profile. The axes are labeled in
      arbitrary units. The vertical lines exemplify typical values for
      $\theta_\mathrm{fit}$ and $\theta_\mathrm{cut}$; see text for
      details.}
    \label{fig:model-cluster+filament}
\end{figure}

Fig.~\ref{fig:model-cluster+filament} shows a simple one-dimensional
toy model of the mass distribution of a cluster with a filamentary
extension. The model consists of the following components: We assume a
cluster with a King profile. This is the solid line in
Fig.~\ref{fig:model-cluster+filament}. In all simulations we see that
the clusters are not spherical but triaxial with their major axes
oriented approximately towards each other. We account for this in the
model by stretching the right half of the King profile (long dashed
line) by a factor $f$, which has to be determined, i.e. the original
profile $\kappa_\mathrm{King}(\theta)$ is replaced with
$\kappa_\mathrm{stretch}(\theta) \equiv
\kappa_\mathrm{King}(f\theta),\, 0<f\le1$, for positive values of
$\theta$. We will call $f$ the ``stretch factor''. The contribution
of the filament (dotted line) $\kappa_\mathrm{fil}(\theta)$ is added to
the stretched King profile. The result is the observed surface mass
density profile on the right-hand side (short dashed) which can be
described by
\begin{eqnarray}
  \kappa_\mathrm{obs}(\theta) = \kappa_\mathrm{King}(f\theta) +
  \kappa_\mathrm{fil}(\theta)\; .
\end{eqnarray}

Since the mass profile of the filament alone is not accessible by
observations, we have to determine a point on the observed mass
profile $\kappa_\mathrm{obs}(\theta)$ that we treat as the ``end" of
the cluster and the ``start" of the filament. We tried this using the
following procedure: The unstretched King profile, observed on the
left-hand side where $\theta < 0$, is stretched by the factor $f$, to
model the influence of tidal stretching. By this step we try to
obtain the (unobservable) cluster profile
$\kappa_\mathrm{stretch}(\theta)$ on the right side without the
contribution of the filament.

This stretched profile is then compared to the observed profile
$\kappa_\mathrm{obs}$ containing the contribution of the filament by
computing the goodness of fit 
\begin{eqnarray}
  \label{eq:4}
  \chi^2 = \sum_{i=i_0}^N
  \left(\frac{\kappa_\mathrm{stretch}(\theta_i) - 
      \kappa_\mathrm{obs}(\theta_i)}{\sigma_i}\right)^2~,
\end{eqnarray} 
at sample points $\theta_i$ in the reconstruction along the main axis
of the system. Typically, the spacing of the sample points will be
that of the grid on which the reconstruction was performed. Linear
interpolation between grid points will be used if the sampling points
do not exactly coincide with the grid points. $\sigma_i$ is the
estimated error in $\kappa_\mathrm{obs}$ at the $i$th point. We define
our sample points such that $\theta_0 = 0$, i.e. $\theta_0$ is placed
at the cluster center. Usually, we will set $i_0 = 1$. Note that in our
model the contribution from the point $\theta = 0$ always vanishes as
by definition the observed and the stretched profile have the same
value.

$\chi^2$ is repeatedly computed for increasing values of $N$. We can
define the ``end of the cluster'' and the ``start of the filament'' by
the point $\theta_\mathrm{cut} \equiv \theta_N$, where the probability
that $\kappa_\mathrm{stretch}$ is a good representation of
$\kappa_\mathrm{obs}$ falls below a pre-defined level, which we call
the ``cut-off confidence level'' $\chi_\mathrm{cut}^2$.

We now have to find a way to determine the stretch factor $f$. For
this, we assume that a position $\theta_\mathrm{fit},\; 0 \le \theta
\le \theta_\mathrm{fit} < \theta_\mathrm{cut}$ exists, such that the
influence of $\kappa_\mathrm{fil}(\theta<\theta_\mathrm{fit})$ is
negligible, i.e. we assume that the observed profile is a fair
representation of the (unobservable) stretched profile
$\kappa_\mathrm{stretch}$. The stretch factor $f$ can then be
determined by fitting the unstretched profile, which we obtain from
observations at $\theta<0$, to the inner portion (i.e. at $\theta \le
\theta_\mathrm{fit}$) of the observed profile. This ``stretch factor
fit'' was done using a $\chi^2$ minimization.

The ``cut-off parameter'' $\theta_\mathrm{cut}$ and the ``cut-off
confidence level'' $\chi_\mathrm{cut}^2$ have to be determined from
simulations.
\begin{figure}
  \resizebox{\hsize}{!}{\includegraphics{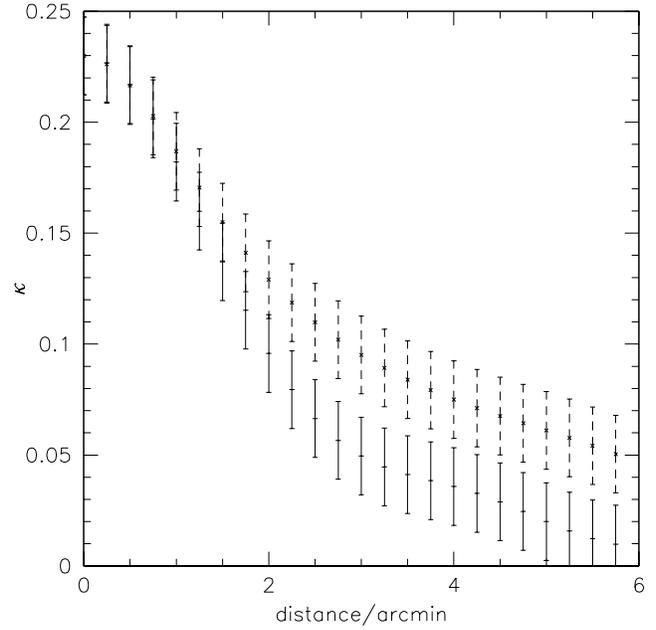}}
  \caption{Surface mass density profiles of the cluster on the left
    in the reconstruction displayed in
    Fig.~\ref{fig:rec.48.cluster_4.4} along the line connecting both
    cluster centers. The crosses mark the surface mass density in the
    filament part, the dashes the surface mass density on the lefthand
    side of the cluster. The $x$-axis denotes the distance from the
    cluster center in arcminutes.}
    \label{fig:chi2-method-sim}
\end{figure} 
Fig.\ref{fig:chi2-method-sim} shows the mass profiles to the left
and right of the center of the left cluster in the
reconstruction displayed in Fig.~\ref{fig:rec.48.cluster_4.4} along
the main axis of that system. For simplicity the error bars were
assumed to be equal to the standard deviation of a reconstructed mass
map of a randomized catalog of background galaxies.

We determined several combinations of the cut-off parameter
$\theta_\mathrm{cut}$ and confidence level $\chi_\mathrm{cut}^2$ that
match the visual impression of filament beginning and cluster end.
However, if these were applied to clusters from other simulations, the
separation point between cluster and filament was placed at
non-sensical positions.

We also modified the starting position $\theta_{i_0}$ in the summation
in eq.~(\ref{eq:4}). First, we placed it at $\theta_\mathrm{fil}$ in
order to exclude the central region, which by definition of this
procedure has a small $\chi^2$. Second, we calculated $\chi^2$ in a
moving window of fixed size and set the separation point between
cluster and filament to the start of the window for which $\chi^2$
fell below the cut-off confidence level. This was done for various
window sizes and confidence levels. Again, parameters that worked
well for one cluster failed completely for others in both approaches.

\subsection{Quadrupole moment map of A~222/223}
\label{sec:quadr-moment-maps}
\begin{figure*}
  \sidecaption
  \includegraphics[width=12cm]{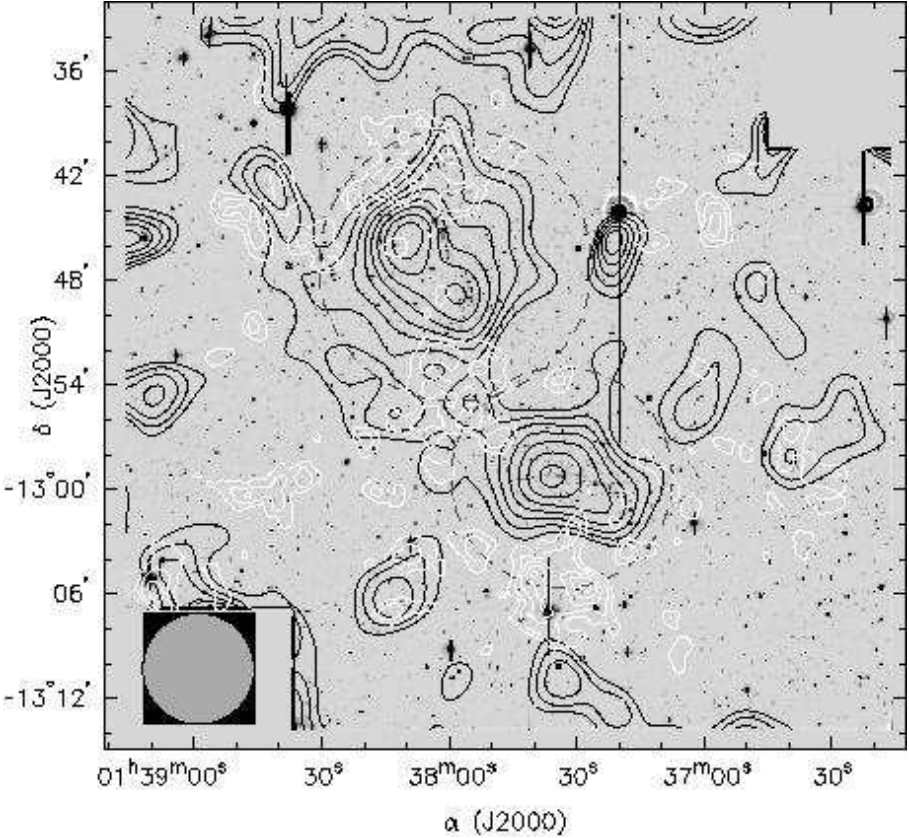}
  \caption{Aperture quadrupole moment map of A~222/223 in an aperture
    of $3\farcm{2}$ radius, corresponding to the radius of the
      circle at the lower left corner. The thick white lines are SNR
    contours for $|Q^{(2)}|$, the lowest contour being at $2$ and
    higher contours increasing in steps of $0.5$. The aperture
    quadrupole moment on the "filament" region reaches a peak SNR of
    $3.0$. The black lines are the contours of the mass
    reconstruction. The dashed circles indicate the $r_{200}$ from our
    best fit NFW models.}
  \label{fig:a222+3.qpole}
\end{figure*}
Having found in the previous section that aperture quadrupole moments
in principle can be used to quantify the presence of a filamentary
structure, if we are able to choose the right size of the aperture, we
now apply this method to the A~222/223 system.
Fig.~\ref{fig:a222+3.qpole} shows a $|Q^{(2)}|$ map with a weight
function with $3\farcm{2}$ radius. The white significance contours
show a quadrupole moment signal on the filament that reaches a peak
SNR of $3.0$. The filter scale was chosen to be the same in which the
aperture mass statistics gave the most significant signal in the
intercluster region. The $|Q^{(2)}|$ signal does not fully trace the
filament candidate but only the Western part of it and an extension
towards the mass peak in the East. The peak SNR is most likely
enhanced by the trough at the Western edge of the mass bridge. It is
not surprising that the significance of the quadrupole moments on the
possible filament is not very high. The aperture mass statistics
already gave a relatively low SNR. The AMM statistics uses data within
the same aperture as $\map$ but gives more information, namely instead
of the mass in the aperture, it gives the mass distribution. Being
generated from the same information, this naturally comes with a lower
SNR.

Several other features -- most of them associated with the slopes of the
two massive clusters -- are also seen in Fig.~\ref{fig:a222+3.qpole}.
Interesting in the context of quantifying filaments is the $|Q^{(2)}|$
statistics in the region between A~222 and the newly detected cluster
SE of it. Indeed we see a signal with a peak signal-to-noise ratio of
$2.7$ extending between the two clusters. The mass reconstruction in
Fig.~\ref{fig:A222+3.rec} also shows a connection between both
clusters but at a level that is dominated by the noise of the
$\kappa$-map. The redshift difference between A~222 and the new
cluster, inferred from the color-magnitude diagram
(Fig.~\ref{fig:blob.rcs}), makes it very unlikely that these two
clusters are connected by a filament. It is much more probable that we
are in the situation depicted by Fig.~\ref{fig:qpole_illus_fake} in
which the influence of the two individual clusters leads to a
quadrupole moment in the intercluster region.

\section{Discussion and conclusions}
\label{sec:discusssion}
Based on observations made with WFI at the ESO/MPG 2.2~m telescope we
find a clear lensing signal from the Abell clusters A~222 and A~223.
Comparing our lensing analysis with the virial masses and X-ray
luminosities, we find that A~222/223 forms a very complex system. Mass
estimates vary considerably depending on the method. Assuming the
best-fit NFW profiles of Sect.~\ref{sec:weak-lens-reconstr}
$M_{200}(\mathrm{A~222}) = 3.0^{+0.7}_{-0.8} \times 10^{14}
h_{70}^{-1} M_{\sun}$ and $M_{200}(\mathrm{A~223}) = 5.3^{+1.6}_{-1.4}
\times 10^{14} h_{70}^{-1} M_{\sun}$. The masses of the best-fit SIS
models within $r_{200}$ as determined from the NFW fit are higher for
both cluster but compatible within their respective error bars:
$M_\mathrm{SIS}(\mathrm{A~222}) = 4.8^{+1.0}_{-1.1} \times 10^{14}
h_{70}^{-1} M_{\sun}$ and $M_\mathrm{SIS}(\mathrm{A~223}) =
7.3^{+1.3}_{-1.4} \times 10^{14} h_{70}^{-1} M_{\sun}$. These mass
estimates are considerably lower than those derived from the virial
theorem for an SIS model. Using the velocity dispersions from
\citetalias{2002A&A...394..395D} we find
$M_\mathrm{vir}(\mathrm{A~222}) = 9.7^{+1.9}_{-1.6} \times 10^{14}
h_{70}^{-1} M_{\sun}$ and $M_\mathrm{vir}(\mathrm{A~223}) =
12.0^{+2.5}_{-2.1} \times 10^{14} h_{70}^{-1} M_{\sun}$.

The $M/L$ ratios we found in Sect.~\ref{sec:light-distr-222223} are
lower than the ones determined by \citetalias{2002A&A...394..395D} of
$M/L_R = (202 \pm 43 )\,h_{70}\,M_{\sun}/L_{\sun}$ for A~222 and
$M/L_R = (149 \pm 33 )\,h_{70}\,M_{\sun}/L_{\sun}$ within a radius of
$1.4\,h^{-1}$~Mpc but agree within the error bars of our values for
the $M/L$ ratios determined from the SIS model masses, and in
  the case of A~223 also with the $M/L$ ratio from the NFW model. Two
competing effects are responsible for this difference. First and
foremost, the weak lensing masses are lower than the masses
\citetalias{2002A&A...394..395D} used. Second, also the luminosities
determined are lower than in \citetalias{2002A&A...394..395D}. This
has two reasons. First, \citetalias{2002A&A...394..395D} analyze the
Schechter luminosity function; this allows them to estimate the
fraction of the total luminosity they observe, while we limit our
analysis to the actually observed luminosity. Also,
\citetalias{2002A&A...394..395D} correct the area available to fainter
objects by subtracting the area occupied by brighter galaxies, which
might obscure fainter ones. Both differences mean that we probably
underestimate the total luminosity of the clusters. Our $M/L$ ratios
are already at the lower end of common $M/L$ ratios. A higher
luminosity would lead to even lower $M/L$ value making A~222 and A~223
unusually luminous clusters, considering their mass. Although this
system is complex and probably still in the process of collapsing, the
$M/L$ ratios of both clusters are very similar and do not exhibit
variations like those observed by \citetalias{2002ApJ...568..141G} in
A~901/902. Variations between mass, optical, and X-ray luminosity are
seen on smaller scales in A~223. A~223-N is very weak in the X-ray
image, while it is the dominant sub-clump in the mass and
optical luminosity density map. The latter may, however, be affected
by the color selection that misses many of the unusually blue bright
galaxies in A~223, especially in the Southern sub-clump.

The weak lensing mass determination depends on the redshift of the FBG
which we assumed to be $\overline{z}_\mathrm{FBG} = 1$. This
assumption is based on the redshift distribution of the
\citet{1999A&A...343L..19F} HDF-S photometric redshift catalog.
Changes in the redshift distribution could change the absolute mass
scale while leaving the dimensionless surface mass density and hence
also the significance of the weak lensing signal unchanged. However,
the A~222/223 clusters are at comparably low redshift and changes in
$\overline{z}_\mathrm{FBG}$ affect the mass scale only weakly. To
bring $M_\mathrm{SIS}(\mathrm{A~222})$ to the value of
$M_\mathrm{vir}(\mathrm{A~222})$, the mean redshift of the faint
background galaxies would have to move to $\overline{z}_\mathrm{FBG} =
0.3$. This is clearly unrealistic given the depth of our WFI images
and the color selection we made. Although we cannot exclude deviations
from the redshift distribution of \citet{1999A&A...343L..19F}, it is
much more realistic to attribute the differences between virial and
weak lensing masses to intrinsic cluster properties.

From the visual impression of the galaxy distribution it is already
obvious that this system is far from being relaxed. This can affect
the measured masses in several ways: First, the deviation from
circular symmetry certainly implies that the line of sight velocity
dispersion is not equal to the velocity dispersion along other axes in
the clusters. If the clusters are oblate ellipsoid with their major
axis lying along the line of sight, the measured velocity dispersions
will overestimate the velocity dispersion. Second, if the clusters are
not virialized, estimating their masses from the virial theorem of
course can give significant deviations from their actual mass.
Finally, we could only successfully obtain weak lensing mass estimates
with spherical models, which are probably not a good representation of
the actual system. Although both clusters are clearly elliptical, fits
with SIE models could not reliably reproduce the observed cluster
properties. This does not come as a total surprise;
\citet{2002A&A...383..118K} already noticed that the shear
log-likelihood function is much more sensitive to changes in the slope
than to a possible cluster ellipticity. The insensitivity of the
log-likelihood function to the ellipticity parameters means that the
fitting procedure rather changes other cluster parameters than
reproducing the actual ellipticity which we see in the parameter-free
weak lensing reconstruction. This behavior and the difficulty to
accurately fit elliptical models to shear data is confirmed by our
simulations in Sect.~\ref{sec:fitt-ellips-galaxy}. Although our
assumption about the dispersion of intrinsic galaxy ellipticities and
number density were much more optimistic than justified by our data,
we could not recover the ellipticity and orientation of the clusters
in the $N$-body simulation.

We found that the concentration parameter $c$ of the NFW profile is
poorly constrained if we omit the central regions of the cluster in
order to avoid contamination with cluster galaxies. This does not
significantly affect the masses determined from fitting NFW profiles
and was not of prime importance to the work presented here. From
varying the radius of the circles in which shear information was
ignored, we saw that galaxies closer to the cluster center constrain
the concentration parameter better than those at large distances from
the clusters. If one wants to determine concentration parameters more
reliably, the fitting procedure has to be extended to include
background galaxies close in projection to the cluster centers, while
ensuring that faint cluster galaxies do not have a strong influence on
the shear signal.

The lensing reconstruction shows a ``bridge'' extending between both
clusters of the double cluster system. We devoted much effort to
developing a method that could objectively decide whether this
tantalizing evidence is indeed caused by a filament like it is
predicted from $N$-body simulations of structure formation.
Unfortunately, this was mostly done without success. The aperture
quadrupole moment statistics in principle has the power to detect the
presence of a filament-shaped structure. To objectively apply it, one
however needs to be able to separate clusters from the filaments
connecting them. We did not find an objective way to do this and had
to resort to subjectively defining the sizes of the apertures used.

We would like to stress that this is not a problem of the weak lensing
technique but stems from the fact that the description of the cosmic
web as filaments and galaxy clusters is based on the visual impression
of $N$-body simulations. Attempts to objectively separate these two
components from each other require a mathematical description which we
tried to develop in Sect.~\ref{sec:defin-clust-filam}. This was mostly
unsuccessful because we could not find a procedure that reliably
reproduces our visual impression. The visual impression of what a
filament is, is often sufficient in simulations or redshift surveys
where filaments stretching long distances between clusters are seen.
In the case of close pairs of clusters -- where we can hope to see
filaments with today's telescopes -- a more objective criterion is
important, but difficult to find.

We have not addressed the question how to distinguish the aperture
quadrupole moment of a filamentary structure from that of a pure
double cluster system in Sect.~\ref{sec:using-apert-mult}. We found
that the quadrupole moment in a system with a filament exceeds that of
a halo-halo system without filament. Closer inspection reveals that
the shape of the quadrupole moment in the intercluster region changes
if a filament is added to a two halo system. Because one can compute
significances for AMM in a limited spatial region, a significant
deviation from the expected shape of the quadrupole moment from a pure
halo system could possibly be used to overcome this difficulty. This
can only work if the signal-to-noise ratio of the aperture quadrupole
moment is high. Possibly stacking several cluster pairs could provide
a sufficiently high SNR. This could in principle be tested with our
$N$-body simulations but is beyond the scope of this paper in which we
try to develop a criterion to quantify the evidence for filaments in
single systems, like the A~222/223 system at hand.

What can we then say about a possible filament between A~222/223? All
observations presented in this work -- weak lensing, optical, and
X-ray -- show evidence for a ``filament'' between the two clusters.
The most compelling evidence probably comes from the number density of
color-selected early-type galaxies, which is present at the $7\sigma$
level (Fig.~\ref{fig:nden.lumden}). The spectroscopic work of
\citetalias{2002A&A...394..395D} and \citet{2000A&A...355..443P}
confirmed the presence of at least some galaxies at the cluster
redshift in the intercluster region. Obtaining a larger spectroscopic
sample in the intercluster region would allow us to spectroscopically
confirm the significance of this overdensity and could provide
insights into the correlation of star formation rates and matter
density \citep[e.g.][]{2004MNRAS.347L..73G}. The X-ray emission
between the clusters is aligned with the overdensity in galaxy number
and luminosity density. This provides further evidence for a
``filament'' extending between A~222 and A~223.

The signal level of the possible filament in the weak lensing map
Fig.~\ref{fig:A222+3.rec} is rather low compared to the clusters. The
aperture quadrupole statistics has a signal at the $3\sigma$ level on
the filament candidate but this signal may already be contaminated by
the outskirts of the cluster in the aperture. The most striking
``feature'' of the mass bridge seen in the $\kappa$ map is the
misalignment with respect to the possible filament seen in the optical
and X-ray maps. This can be interpreted in several ways. It could
suggest that the surface mass density on the ``true filament'' defined
by the position of the optical overdensity and X-ray emission is below
our detection limit and what we see in the $\kappa$ map is a noise
artifact. This possibility aside, the observed misalignment can have
several causes. First, as we already discussed in
Sect.~\ref{sec:light-distr-222223}, the influence of the many
reflection features around the bright star West of A~223 on the weak
lensing reconstruction is difficult to determine. It seems that the
cluster peaks are shifted preferentially away from the reflection
rings. The same could be true for the ``filament'' in the
reconstruction. Second, the position of structures inferred from weak
lensing is affected by the noise of the reconstruction. This is
especially true for low mass structures and is illustrated by our
simulations using SIS models to infer the positional uncertainty of
weak lensing reconstructed peaks in
Sect.~\ref{sec:weak-lens-reconstr}. It is possible that at least part
of the observed misalignment is caused by the noise of the weak
lensing method. Finally, one could in principle imagine that the
off-set is real and a misalignment of dark and luminous matter is
present. This would require complex and possibly exotic physical
processes that cause galaxies to form next to a dark matter filament
and not in it. At present there is no good observational support for
such a scenario. We should note, however, that a misalignment between
mass and light is also present in the filament candidate of
\citetalias{2002ApJ...568..141G}.

As we have not found an objective way to define what a filament in a
close double cluster pair is, the question whether what we observe in
A~222/223 constitutes a filament or not can also not be answered
objectively. Thus, our filament candidate is -- in this respect -- not
very different from those of \citet{1998astro-ph/9809268K} and
\citetalias{2002ApJ...568..141G}. The \emph{projected} virial radii of
the clusters marginally overlap. However, (1) the redshift difference
between the clusters make an actual overlap of the clusters unlikely;
(2) the projected mass in clusters falls off steeply, and weak lensing
is currently not capable of mapping the cluster mass distribution out
to the virial radius. A signature of a filament should thus already be
present inside the virial radius.

The unambiguous weak lensing detection of a filament between two
clusters would provide a powerful support for the theory of structure
formation and the ``cosmic web''. Taking the $3\sigma$ signal of the
quadrupole statistics on the filament candidate at face value, an
increase of the number of density of FBGs by a factor of $2.8$ could
give a $5\sigma$ detection. Such number densities can be reached by
$8$~m class telescopes. The A~222/223 system, being only the third
known candidate system to host a filament connecting two cluster, would
be a good target for such a study. In fact, a weak lensing study of
A~222/223 using SuprimeCam at the Subaru telescope is already underway
(Miyazaki et al., in preparation).

In addition to the lensing signal from Abell~222 and Abell~223 we
found a significant mass peak SE of A~222. This peak coincides with an
overdensity of galaxies. The color-magnitude diagram of these galaxies
suggest that this newly found cluster is at a redshift of $z\sim0.4$,
but this estimate comes with a considerable uncertainty and requires
spectroscopic confirmation. A maximum likelihood fit to the shear data
around this mass peak leads to a best-fit SIS model with a velocity
dispersion of $728^{+101}_{-120}$~\kms. This serendipitous detection
again illustrates the power of weak lensing as a tool for cluster
searches.

\begin{acknowledgement}
  We wish to thank the anonymous referee for many comments that helped
  to improve this paper. This work has been supported by the German
  Ministry for Science and Education (BMBF) through DESY under the
  project 05AE2PDA/8, and by the Deutsche Forschungsgemeinschaft under
  the project SCHN 342/3--1.
\end{acknowledgement}

\appendix
\section{Strong lensing features in A~222}
\label{sec:strong-lens-feat}
\begin{figure}
  \begin{center}
    \resizebox{\hsize}{!}{\includegraphics{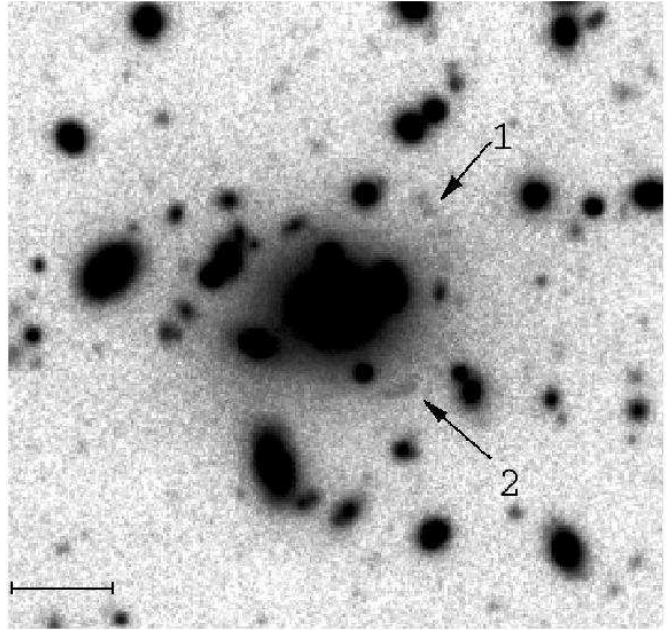}}
    \caption{Arclet candidates around the cD galaxy in A~222. North is
      up and East is to the left. The scale is 10\arcsec~long.}
    \label{fig:a222.arcs.arrows.inv}
  \end{center}
\end{figure}

Already in 1991
\citet[][SEF]{1991MNRAS.252...19S}\defcitealias{1991MNRAS.252...19S}{SEF}
found two candidate arclets in the center of A~222. We also see two
possible arclets in the center of A~222 displayed in
Fig.~\ref{fig:a222.arcs.arrows.inv}.
Arclet 1 is the same as found by \citetalias{1991MNRAS.252...19S} and
labeled A~222-1. Unfortunately, \citetalias{1991MNRAS.252...19S}'s
second candidate is not marked on the plate in their paper, and as
\citetalias{1991MNRAS.252...19S} give only distances from the cluster
center and no position angle we do not know whether their second
candidate corresponds to ours. A comparison of the arclet candidate
properties between \citetalias{1991MNRAS.252...19S} and our candidates
is given in table~\ref{tab:arclets}.
\begin{table}[htbp]
  \begin{center}
    \caption{Arclet candidate properties from \citetalias{1991MNRAS.252...19S} and our data. The
      column entries are distance from the center of the cD galaxy,
      axis ratio, and position angle measured clockwise from the north
      direction in Fig.~\ref{fig:a222.arcs.arrows.inv}.}
    \begin{tabular}{lrrr}
      \hline
      Arc ID & d/arcsec & b/a & pos. angle\\\hline\hline
      \multicolumn{4}{l}{\textbf{\citetalias{1991MNRAS.252...19S}}}\\
      A 222-1 & 12.2 & 4.7 & - \\
      A 222-2 & 14.1 & 2.6 & - \\
      \multicolumn{4}{l}{\textbf{this work}}\\
      arclet 1 & 12.7 & 2.8 & 42\degr \\
      arclet 2 & 10.8 & 3.0 & 140\degr \\\hline
    \end{tabular}
    \label{tab:arclets}
  \end{center}
\end{table}
The distance measurements for A~222-1 and arclet 1 are in good
agreement but the values for the axis ratio show a clear deviation.
The difference may be due to the comparably poor image quality in the
work of \citetalias{1991MNRAS.252...19S} and blending with the nearby
object to the South-West of arclet 1. However, it must also be
mentioned that the determination of the axial ratio is relatively
uncertain and we estimate its error to be of the order $\simeq 0.6$.

Given the discrepancy between the distance measurements for A~222-2
and arclet 2 it is unlikely that these are the same objects.

Unfortunately, the $V$ band image is not deep enough to show the
candidate arclets, so that no color information is available.

\section{Multipole moments}
\label{sec:multipole-moments}
\citet{1997MNRAS.286..696S} define the complex $n$th-order aperture
multipole moment as
\begin{eqnarray}
  \label{eq:5}
  Q^{(n)}\left( \vec{\theta}_0 \right) = 
  \int_0^\infty \drm^2 \vec{\theta}\; \theta^n U(|\vec{\theta}|) 
  \e^{n\im\varphi}\kappa(\vec{\theta}_0 +
  \vec{\theta}) 
\end{eqnarray}
with a radially symmetric weight function $U(|\vec{\theta}|)$. For
$n=2$ eq.~(\ref{eq:5}) expresses the aperture quadrupole moment in
terms of the surface mass density. Based on this definition, an
expression for the aperture moments in terms of shear estimates may be
found \citep{1997MNRAS.286..696S}:
\begin{eqnarray}
  \label{eq:6}
  Q^{(n)}\left( \vec{\theta}_0 \right) & = & 
  \frac{1}{\overline{n}} 
  \sum_{i=1}^{N} \e^{n\im\varphi_i} \times \nonumber\\
  & & \left\{
    \theta_i^n U(\theta_i) \varepsilon_{\mathrm{t}i} +
    \im \frac{\theta_i^n [
      nU(\theta_i) + \theta U'(\theta_i)
      ]}{n} \varepsilon_{\times i}
    \right\}\;,
\end{eqnarray}
where $\overline{n}$ is the number density of galaxies in the
aperture, $(\theta_i, \varphi_i)$ are the polar coordinates of the
$i$th galaxy with respect to $\vec{\theta}_0$, and
$\varepsilon_{\mathrm{t}i} = - \mathcal{R}(\varepsilon_i
\e^{-2\im\varphi_i})$ and $\varepsilon_{\times i} = -
\mathcal{I}(\varepsilon_i \e^{-2\im\varphi_i})$ are the tangential and
cross components of the shear estimate, respectively, with respect to
$\vec{\theta}_0$. Here $U'(\theta)$ is the derivative of the weight
function.

We now show that the definition (\ref{eq:5}) cannot be generalized to
non-radially symmetric filters $U(\vec{\theta})$. We partially
integrate eq.~(\ref{eq:5}) with respect to $\varphi$ and obtain
\begin{eqnarray}
  \label{eq:7}
   Q^{(n)} & = &  
   \frac{\im}{n} \int_0^\infty \drm \theta\; \theta^{n+1}\nonumber\\
   & &\int_0^{2\pi} \drm \varphi\; \e^{n\im\varphi} 
   \left [ U(\vec{\theta}) \frac{\partial}{\partial
       \varphi}\kappa(\vec{\theta}) + \kappa(\vec{\theta})
     \frac{\partial}{\partial\varphi} U(\vec{\theta})\right]\;,
\end{eqnarray}
where for simplicity we have set $\vec{\theta}_0 = 0$ without loss of
generality. The integral over the first term in this expression can
be expressed in terms of the shear in analogy to
\citet{1997MNRAS.286..696S}, while we integrate the second term again
by parts, this time with respect to $\theta$. This integration over
the second term leads to an expression similar to (\ref{eq:7}) with
integrals over two terms; one that can be readily expressed in terms
of the shear, the other requiring further integration by parts, and so
on. Eventually, the aperture multipole moment can be expressed in
terms of the shear as an infinite series of integrals:
\begin{eqnarray}
  \label{eq:8}
  Q^{(n)} & = & 
  \int_0^\infty \drm \theta\; \theta^{n+1} \int_0^{2\pi} \drm \varphi\;
  \e^{n\im\varphi} U(\vec{\theta}) 
  \gamma_\mathrm{t}(\vec{\theta})\nonumber\\
  & & +\frac{\im}{1-\frac{\im}{n}} \int_0^\infty \drm \theta \int_0^{2\pi}
  \drm \varphi\; \theta^{n+1} \e^{n\im\varphi} U(\vec{\theta}) 
  \gamma_{\times}(\vec{\theta})\nonumber\\
  & &+\sum_{j=0}^\infty \left(\frac{\im}{n}\right)^{j+1} 
  \!\!\int_0^\infty\drm \theta \int_0^{2\pi}\drm \varphi\; \theta^{n+2} 
  \gamma_{\times}(\vec{\theta}) \frac{\partial^{j+1}}{\partial\theta
    \partial\varphi^j} U(\vec{\theta})
\end{eqnarray}
with the requirement on the weight function that
\begin{eqnarray}
  \label{eq:9}
  \gamma_{\times}\theta^{n+2}\left(\frac{\partial}{\partial\varphi}\right)^j 
  U(\vec{\theta}) \rightarrow 0,\; \mathrm{for}\; \theta \rightarrow 0\;
  \mathrm{and}\; \theta \rightarrow \infty\;,
\end{eqnarray}
so that the integrals exist. $\gamma_\mathrm{t}$ and $\gamma_{\times}$
are the tangential and cross components of the shear, in analogy to
$\varepsilon_\mathrm{t}$ and $\varepsilon_{\times}$ above. It turns out
that this sum in general does not converge. E.g. for $n=2$ and
\begin{eqnarray}
  \label{eq:10}
  U(\vec{\theta}) = f(\theta)\left[ 1+\frac{\epsilon}{2}
    \left(\e^{2\im\varphi} + \e^{-2\im\varphi}\right)\right]\;,
\end{eqnarray}
where $f(\theta)$ is a smooth, positive, and finite function
satisfying the condition~(\ref{eq:9}), the sum in eq.~(\ref{eq:8})
oscillates around $0$. This behavior can be understood if we insert
(\ref{eq:10}) into (\ref{eq:5}) and integrate by parts, while setting
$\kappa(\vec{\theta})$ to a constant value $\kappa_0$. The quadrupole
moment then depends on $\kappa_0$, unless we allow $\theta^3
f(\theta)$ to be compensated. A constant $\kappa_0$ does not influence
the shear. Hence, the sum in eq.~(\ref{eq:8}) cannot converge. We thus
find that the mass-sheet degeneracy prevents us from computing
aperture multipole moments in non-circular apertures.

\bibliographystyle{aa}
\bibliography{1523}

\end{document}